\documentclass[12pt]{article}
\usepackage{a4wide, amsmath, latexsym, epsfig,
  graphicx, rotating, fancyhdr, tabularx, afterpage}
\usepackage{subfigure}

\newcommand\ba{\begin{eqnarray}}
\newcommand\ea{\end{eqnarray}}

\begin{document}
\begin{titlepage}

\begin{flushright}
{ IFJPAN-IV-2011-14 \\ CERN-PH-TH/2011-309}
\end{flushright}

\vspace{0.2cm}
\begin{center}
{\Huge \bf Bremsstrahlung simulation in   $K \to \pi l^\pm\nu_l (\gamma)$ decays}
\end{center}

\vspace*{5mm}

\begin{center}
   {\bf    Qingjun Xu$^{a,b}$ and Z. W\c{a}s$^{b,c}$}\\
{\em $^a$ Department of Physics, Hangzhou Normal University,
Hangzhou 310036, China  }\\
       {\em $^b$  Institute of Nuclear Physics, PAN,
        Krak\'ow, ul. Radzikowskiego 152, Poland}\\
{\em $^c$ CERN PH-TH, CH-1211 Geneva 23, Switzerland }
\end{center}
\vspace{.1 cm}
\begin{center}
{\bf   ABSTRACT  }
\end{center}
In physics simulation chains, the
{\tt PHOTOS} Monte Carlo program is often used  to simulate
 QED effects in decays
of intermediate particles and resonances.

The program is based on an exact
multiphoton phase space.  In general, the matrix element  is
obtained from
 iterations of a universal kernel and approximations are involved.
To evaluate the program  precision, it is necessary
to formulate and implement within the generator the exact matrix element,
which depends on the  decay channel.
Then,  all terms necessary for non-leading
logarithms are taken into account.
 In the present letter we  focus
on the decay $K\to\pi l^\pm  \nu_l$ and tests of the {\tt PHOTOS}
Monte Carlo program.   We conclude a  0.2\% relative precision
 in the implementation
of the hard photon  matrix element  into the
emission kernel, including the case where approximations are used.

\centerline{{\it Submitted to Eur. Phys. J.} C }

 \vspace{1cm}
\begin{flushleft}
{   IFJPAN-IV-2011-14 \\  CERN-PH-TH/2011-309\\
 December, 2011}
\end{flushleft}

\vspace*{1mm}
\bigskip
\footnoterule
\noindent
{\footnotesize \noindent  $^{\dag}$
 Work of ZW is supported in part by the
Polish Ministry of Science and Higher
Education grant No. 1289/B/H03/2009/37., Q. Xu is supported by China-Poland
inter-governmental cooperation grant 34-13,
National Natural Science Foundation of China under Grant No. 11147023
and Zhenjiang Provincial Natural Science Foundation of China under Grant No. LQ12A05003.}
\end{titlepage}

\section {Introduction}

Semileptonic flavour changing decays, such as $K \to \pi l^\pm
\nu_l$ offer a window for measurements of Standard Model
basic couplings: quark mixing angles~\cite{Antonelli:2010yf}.
Moreover properties of low and medium energy hadronic interactions
manifest themselves in such decays. It is thus important to keep
control of decay products
distributions in a form suitable for comparisons with data and
without loosing control of the underlying quark level matrix element.

Comparison between experimental data and theoretical predictions relies
 on Monte Carlo simulation to take into
account the  detector
response \cite{Actis:2010gg}. Given today's experimental precision,
generators used in such comparisons must be based on exact phase space and explicit formulation
of the matrix
element. Otherwise the discussion of theoretical uncertainties
in realistic applications is rather difficult.

QED bremsstrahlung must be
taken into account in these comparisons too.
Infrared singularities cancel out
in sufficiently inclusive  observables.  In a first approximation, the
QED bremsstrahlung amplitude can be
factorized as the  Born amplitude and an emission (eikonal or collinear) term.
This can be done for a calculation at a  fixed order of perturbation expansion, but
it holds to all orders and is known under the name of  exponentiation \cite{yfs:1961} (for application in Monte Carlo simulation
 see e.g. \cite{yfs1:1988,kkcpc:1999}).
In phase space regions where photons are collinear to charged
particles, factorization theorems define the dominant terms of the amplitudes.
This is why,
bremsstrahlung can be treated to a good precision independently of the
decay channel. As a consequence
predictions, which neglect  QED effects, represent a valid segment
 of the phenomenology work.

For the decays of a given particle,
matrix elements based Monte Carlo generators  are prepared either
by theorists working on  effective lagrangians (also on QCD based predictions),
or by experimental  physicists.
 Since many years the
  {\tt PHOTOS} Monte Carlo program \cite{Barberio:1990ms,Barberio:1994qi} is used
for simulation  of bremsstrahlung in decays.
It represents a separate segment of the simulation chain.

 With the increased precision of
new available  data, such an implementation requires a careful discussion
of its systematic errors, which has to be repeated
for each decay mode. The
phenomenological importance of approximations needs to be analyzed,
and the
factorization of QED emission terms need to be reviewed, using as a reference
solutions based on exact  phase space and  matrix element
for the whole decay (thus including QED bremsstrahlung).

The  {\tt PHOTOS} Monte Carlo program was presented for the fist time in Ref. \cite{Barberio:1990ms}.
The universal
kernel was introduced and was shown to work, within expected accuracy, in cases where
 comparisons with first order matrix element reference simulation programs were available. In references
\cite{Barberio:1994qi} and \cite{Golonka:2005pn}
the notion of iteration was introduced in the  {\tt PHOTOS} program,
 first, for
double photon emission, and later for multiple photon emission.
Because of a rather unique order of iteration (iteration over sources providing
collinear singularities is performed first,  then construction  of consecutive
photon is started), the  algorithm is compatible at the same time with an
exclusive exponentiation and resummation of collinear logarithms.

In  reference \cite{Golonka:2006tw} a discussion of the
exact matrix element implemented within the {\tt PHOTOS}  kernel was performed for the
$Z$ decay.  The discussion was
continued in Ref. \cite{Nanava:2006vv} for the decay of a scalar into pair of scalars. In this study, a detailed
presentation of the phase space parametrization was given and
it was followed in \cite{Nanava:2009vg} by a discussion of the matrix
elements of the  $W$ leptonic decay and the decay of virtual photons  to
pairs of scalars. Matrix element weights
became available for  public use with  \cite{Davidson:2010rw}.

Until now, discussions of matrix elements were addressing
two-body decays only. In this case, not only phase space at Born level is
particularly simple, but also photon emission  is easy to handle.

In the present work, we  focus on the $K_{l3}$  decay for which  a
three-body kinematic is present already at the
Born level. By studying this decay, we  test not only  the matrix element implementation effects,
 but also the {\tt PHOTOS} phase space generation\footnote{Parametrization of phase space in
 {\tt PHOTOS} is explicit and exact,
if the presampler for collinear emissions is  used only along a single charge.
Otherwise, starting from the moment when the  photon is supposed to be
added to a more than two-body configuration, an approximation appears
when the phase space Jacobians for the multitude of branches are
combined. This can be improved, but for test cases with second order matrix
elements in $Z$ decay \cite{Golonka:2006tw}, we have found
 that this approximation is necessary, unless complete double photon emission amplitude would
be installed into
the program at the same time.
}.

Our study is organized as follows. In Section 2 we present the  matrix elements for
$K_{l3}$ decays, at Born level and in the case of single photon
emission. Results of calculations  based on  scalar
QED \cite{Quantum-PS1995} and on ChPT
\cite{Weinberg:1978kz,Gasser:1983yg,Leutwyler:1993iq} with
truncations as in Refs. \cite{Cirigliano:2008wn,Cirigliano:2007zz}
are compared and discussed. In this context we also investigate matrix elements and
their factorization properties. In Section
3, we consider fully differential distributions.
For that purpose squared
 matrix element are reviewed and possible options
resulting from physics assumptions (scalar QED or  \cite{Cirigliano:2008wn,Cirigliano:2007zz})  are shown.
We address again factorization properties, this time stressing
features which are useful when constructing a Monte Carlo program. The  set-up and implementations
prepared for tests are
 presented in Section 4.
Numerical results are collected in Section 5. Conclusions, including an estimate of  {\tt PHOTOS}
Monte Carlo simulation precision
for  $K_{l3}$ decays,  are given in Section 6.

\section {Amplitudes for Ke3$(\gamma)$ decays.}
In the first sub-section we  present the amplitude for the decay of charged kaons. In the second one
the case of the neutral kaon is considered, similarities of the two cases
are investigated as well.
\subsection{ Amplitude for  $K^-(p)\to\pi^0(q)
 + e^-(p_e) + \bar \nu_e(p_\nu)$  decay}
Let us start the discussion from the amplitude of the  decay
\ba
 K^-(p)\to\pi^0(q)+e^- (p_e)+\bar\nu_e(p_\nu) \,
 \label{ke3g1}
\ea taken at the Born level. With the   notation of Ref.
\cite{Cirigliano:2008wn,Cirigliano:2007zz} it reads \ba M^c_{\rm
Born} &=& \frac{e^2V_{us} F_{K\pi}(t)}{8 \sqrt 2 s_W^2}
                    \frac{(p +q)^\mu}{t-M_W^2}\nonumber \\
&& \overline {u}(p_e, \lambda_e)\gamma_\mu(1-\gamma_5)v(p_\nu,
\lambda_\nu)\, , \label{Born0} \ea where $\lambda_e (\lambda_\nu) $
denote  the electron (neutrino) helicity, $F_{K\pi}(t)$ is the form
factor and $t = (p-q)^2$. Because of the relatively small mass of
the $K$ meson,
  $t$ is always $\ll M_W^2$ and the amplitude
simplifies to
 \ba
M^c_{\rm Born} &=& \frac{G_F V_{us} F_{K\pi}(t)}{2} (p + q)^\mu
\nonumber \\
&& \overline {u}(p_e, \lambda_e) \gamma_\mu(1-\gamma_5)v(p_\nu,
\lambda_\nu)\, . \label{Born1} \ea
\begin{figure}[htp!]
\begin{tabular}{ll}
{\includegraphics[%
 width=0.40\columnwidth]{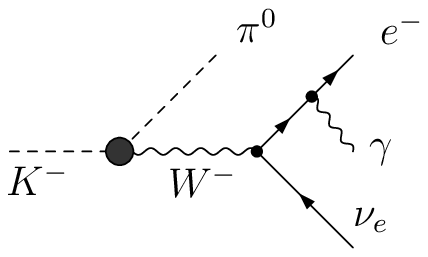}} & {\includegraphics[%
  width=0.40\columnwidth]{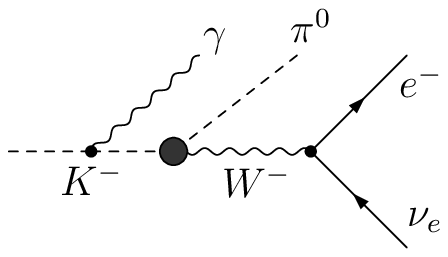}} \\
\ \ \ \ \ \ \ \ \ \ (a) & \ \ \ \ \ \ \ \ \ \ (b)\\
 {\includegraphics[%
  width=0.40\columnwidth]{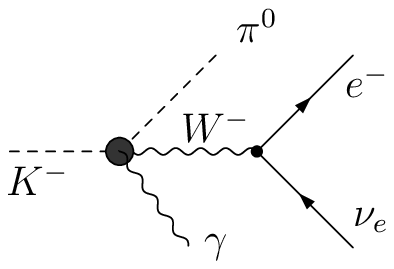}} & {\includegraphics[%
  width=0.40\columnwidth]{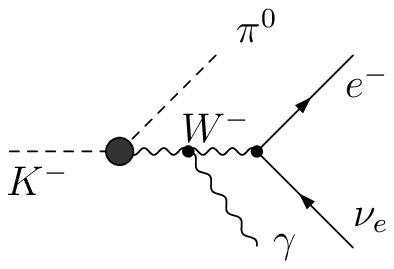}}\\
 \ \ \ \ \ \ \ \ \ \ (c) & \ \ \ \ \ \ \ \ \ \ (d)
\end{tabular}
\caption{Feynman diagrams of the $K^- \to\pi^0 e^- \bar \nu_e
\gamma$ decay.}
\end{figure}
Let us now consider the  amplitudes for single photon emission.
We  follow  Ref. \cite{Nanava:2009vg}  and we  define the
photon polarization states in  the rest frame of the $K$ meson. This choice
 fixes the gauge.  However,
our amplitudes are  expressed in a Lorentz and gauge invariant formulation.

Scalar QED diagrams  for $K^-(p)\to\pi^0(q) e^-(p_e) \bar
\nu_e(p_\nu)$ $\gamma(k)$ decay are presented in Figure~1. The
contribution of diagram (d) is proportional to
 $1/M_W^4$ while for other diagrams it is proportional to $1/M_W^2$. For this reason, we can  neglect
 the contribution of  (d).
The  amplitude of the  $K^- \to\pi^0 e^- \bar\nu_e \gamma$ decay
reads: \ba M^c &=& \frac{G_F V_{us} F_{K\pi}(t)}{2}\overline
{u}(p_e, \lambda_e) \left [Q_e (p + q)^\mu \right. \nonumber \\
& & \left.\left(\frac{p_e \cdot \epsilon}{p_e\cdot k}+
 \frac{\not \epsilon \not k }{2 p_e\cdot k} \right ) - Q_K (p + q-k)^\mu \frac{p \cdot \epsilon}{p\cdot k}
\right. \nonumber \\
& & \left. -Q_K \epsilon^\mu \right]\gamma_\mu (1-\gamma_5)v(p_\nu,
\lambda_\nu)\, , \label{ke3ctotal} \ea where $Q_e, Q_K$ denote the
charges of $e^-$ and $K^-$ respectively. To visualize its
factorization properties this amplitude can be expressed as a sum of
three gauge invariant terms:
  \ba
M^c = M^c_I + M^c_{II} + M^c_{III}\, , \label{basics} 
\ea 
where 
\ba
M^c_I &=&  \frac{G_F V_{us} F_{K\pi}(t)}{2} (p + q)^\mu \left (Q_e
\frac{p_e \cdot \epsilon}{p_e\cdot k}
- Q_K\frac{p \cdot \epsilon}{p\cdot k}\right )\nonumber \\
& & \overline {u}(p_e, \lambda_e) \gamma_\mu
(1-\gamma_5)v(p_\nu, \lambda_\nu)\, , \label{basicsI}\\
M^c_{II} &=& \frac{G_F V_{us} F_{K\pi}(t)}{2} (p + q)^\mu \overline
{u}(p_e, \lambda_e) Q_e\frac{\not \epsilon \not k}{2 p_e\cdot
k}\nonumber \\
& & \gamma_\mu
(1-\gamma_5)v(p_\nu, \lambda_\nu)\, ,\label{basicsII}\\
M^c_{III}&=& \frac{G_F V_{us} F_{K\pi}(t)}{2} Q_K
\left ( k^\mu \frac{p \cdot \epsilon}{p\cdot k} - \epsilon^\mu\right )\nonumber \\
& & \overline {u}(p_e, \lambda_e) \gamma_\mu (1-\gamma_5)v(p_\nu,
\lambda_\nu)\, . \label{basicsIII} \ea The first  term $M^c_I$
consists of the
 Born-level amplitude times an eikonal factor. The second, $M^c_{II}$, is free of soft singularity
but contributes logarithmically in the collinear limit. Finally the third term, $M^c_{III}$,
is free of both soft and collinear singularities. Hence,
formula (\ref{basics}) provides a clearly  structured expression of the amplitude.

For each term (\ref{basicsI},\ref{basicsII},\ref{basicsIII}) a separation into  leptonic
and hadronic parts is visible.  It is encouraging, because  it
coincides with the structure  which was useful in
\cite{Nanava:2009vg}.
In our present work we again see that  the first two parts are process independent in their emission aspect,
 and only the last non-dominant part breaks
this property%
\footnote{Expression (\ref{basicsIII})  coincides also
with the similar term  (\ref{basicsIIIn})  discussed later in the amplitude of
the $K^0 \to \pi^\mp l^\pm \nu_l$ decay.}.
This is why, we expect this
formulation to be useful for our numerical discussion.

To prepare a comparison with  amplitudes  of
Ref.~\cite{Cirigliano:2008wn} the  Born and photon emission
amplitudes (equations~(\ref{Born1}) and (\ref{ke3ctotal})), can be written
thanks to Dirac equation as:
 \ba
 M^c_{\rm
Born} &=& \frac{G_F V_{us} F_{K\pi}(t)}{2}\nonumber \\
&& \overline {u}(p_e, \lambda_e) \left (2 \not q + m_e\right)
(1-\gamma_5)v(p_\nu, \lambda_\nu)\,  \label{Borncnew} \ea and
 \ba
M^c &= &\frac{G_F V_{us} F_{K\pi}(t)}{2} \overline {u}(p_e,
\lambda_e)\left [ \left (Q_e \frac{p_e \cdot \epsilon}{p_e\cdot k} -
Q_K\frac{p \cdot
\epsilon}{p\cdot k}\right )\right. \nonumber \\
&& \left. +Q_e \frac{\not \epsilon \not k }{2 p_e\cdot k}\right ]
\left (2\not q +m_e\right)(1-\gamma_5)v(p_\nu,
\lambda_\nu)\label{ke3ctotalnew}. \hskip 1.5 cm \ea This new
formulation  coincides
 with  formula (13) of
Ref. \cite{Cirigliano:2008wn}.
 In that paper,
 the form factor is taken at $t=0$
(see there,  formulae (2) and (3)), we follow the same choice.

 The new form of
the amplitude can  also be splitted into two gauge invariant parts:
 \ba M^c &= & M_{I^\prime}^c +    M_{II^\prime}^c
\ea 
where 
\ba M_{I^\prime}^c &= & \frac{G_F V_{us} F_{K\pi}(t)}{2}
\overline {u}(p_e, \lambda_e)\left (Q_e\frac{p_e \cdot
\epsilon}{p_e\cdot k} - Q_K\frac{p \cdot \epsilon}{p\cdot k}\right
)\nonumber \\
&& \left (2\not q +m_e\right)(1-\gamma_5)v(p_\nu,
\lambda_\nu)\, ,\label{ke3ctotalnewI} \\
 M_{II^\prime}^c &= & \frac{G_F V_{us}
F_{K\pi}(t)}{2} \overline {u}(p_e, \lambda_e) Q_e\frac{\not \epsilon
\not k}{ 2p_e\cdot k}\nonumber \\
&&\left (2\not q +m_e\right) (1-\gamma_5)v(p_\nu,
\lambda_\nu)\label{ke3ctotalnewII}\, . \ea The first gauge invariant
part has the form of a born-like amplitude times an eikonal factor.
The second one does not contribute to soft singularities, but
contribute to collinear singularities. Unfortunately  formulae
(\ref{ke3ctotalnewI}, \ref{ke3ctotalnewII}) do not manifest the
process independent form which is useful for construction of Monte
Carlo programs.

For all amplitudes presented above as well as their parts, the terms
proportional to the electron mass were carefully kept. The
results naturally extend to  the case of the $K^- \to \pi^0 \mu^-
\bar\nu_\mu$  decay.

\subsection{ Amplitude for  $K^0(p)\to\pi^+(q)
 + e^-(p_e) + \bar\nu_e(p_\nu)$  decay}
The $K^0(p)\to\pi^+(q)
 + e^-(p_e) + \bar\nu_e(p_\nu)$ decay  is
interesting from the point of view of Monte Carlo discussions,
even though in this case effects of bremsstrahlung are of lesser  phenomenological relevance.
 There are not only two charged particles of different masses in the
final state, but there is also  a spectator $\bar\nu_e$,
important  from the point of view of phase space generation.

The charged and neutral $K$ decays amplitudes are quite similar. The
Born level amplitude reads: 
\ba 
M^0_{\rm Born}& =& \frac{G_F V_{us}
F_{K\pi}(t)}{\sqrt 2}
(p + q)^\mu \nonumber \\
&&\overline {u}(p_e, \lambda_e) \gamma_\mu
(1-\gamma_5)v(p_\nu, \lambda_\nu)\, .\label{born-neutral}
\ea

The amplitude for single photon emission is again quite short 
\ba
M^0 &=& \frac{G_F V_{us} F_{K\pi}(t)}{\sqrt 2}\overline {u}(p_e,
\lambda_e) \left [Q_e (p + q)^\mu \right. \nonumber \\
&& \left. \left(\frac{p_e \cdot
\epsilon}{p_e\cdot k}+
 \frac{\not \epsilon \not k }{2 p_e\cdot k} \right )
+ Q_\pi (p + q+k)^\mu \frac{q \cdot \epsilon}{q\cdot k}\right. \nonumber \\
&& \left. -Q_\pi
\epsilon^\mu\right
]\gamma_\mu (1-\gamma_5)v(p_\nu, \lambda_\nu)
\label{ke30ctotal}
\ea
and can be expressed as a sum of three gauge invariant parts:
 \ba
M^0 = M^0_I + M^0_{II} + M^0_{III}\, , \label{no-basic} 
\ea 
where
\ba M^0_I &=& \frac{G_F V_{us} F_{K\pi}(t)}{\sqrt 2} (p + q)^\mu
\left (Q_e\frac{p_e \cdot \epsilon}{p_e\cdot k}
+ Q_\pi\frac{q \cdot \epsilon}{q\cdot k}\right )\nonumber \\
&& \overline {u}(p_e, \lambda_e) \gamma_\mu
(1-\gamma_5)v(p_\nu, \lambda_\nu)\, , \label{basicsIn}\\
M^0_{II} &=& \frac{G_F V_{us} F_{K\pi}(t)}{\sqrt 2} (p + q)^\mu
\overline {u}(p_e, \lambda_e) Q_e\frac{\not \epsilon \not k} {2 p_e\cdot k}\nonumber \\
&& \gamma_\mu
(1-\gamma_5)v(p_\nu, \lambda_\nu)\, ,\\
M^0_{III}&=& \frac{G_F V_{us} F_{K\pi}(t)}{\sqrt 2} Q_\pi\left
(k^\mu \frac{q \cdot \epsilon}{q\cdot k} - \epsilon^\mu \right )\nonumber \\
&&
\overline {u}(p_e, \lambda_e) \gamma_\mu (1-\gamma_5)v(p_\nu,
\lambda_\nu)\, . \label{basicsIIIn}\ea
Here $Q_\pi$ denotes  the $\pi^\pm$ charge. As in the case of $K^\pm$,
only the first part (formula (\ref{basicsIn})) is infrared singular, and the third part is free
of collinear singularity.

With the help of the Dirac equation we can transform (\ref{ke30ctotal}) into:
 \ba M^0 &=& \frac{G_F V_{us}
F_{K\pi}(t)}{\sqrt 2} \overline {u}(p_e, \lambda_e)\left [ \left
(Q_e \frac{p_e \cdot \epsilon}{p_e\cdot k} + Q_\pi\frac{q \cdot
\epsilon}{q\cdot k}\right ) \right. \nonumber \\
&&\left. +Q_e \frac{\not
\epsilon \not k }{2 p_e\cdot k}\right ]\left (2\not q +m_e\right)(1-\gamma_5)v(p_\nu,
\lambda_\nu)\nonumber \\
&& + 2\frac{G_F V_{us} F_{K\pi}(t)}{\sqrt 2} Q_\pi
\left ( k^\mu \frac{q \cdot \epsilon}{q\cdot k} - \epsilon^\mu\right )\nonumber \\
&& \overline {u}(p_e, \lambda_e) \gamma_\mu (1-\gamma_5)v(p_\nu,
\lambda_\nu) \label{ke3neutralnew}, \ea which coincides
with formula (14) of Ref. \cite{Cirigliano:2008wn} though their form are different.
If only the first term of our formula (\ref{ke3neutralnew})
would be taken,
 \ba M_{I^\prime}^0 &=& \frac{G_F V_{us}
F_{K\pi}(t)}{\sqrt 2} \overline {u}(p_e, \lambda_e)\left [ \left
(Q_e \frac{p_e \cdot \epsilon}{p_e\cdot k} + Q_\pi\frac{q \cdot
\epsilon}{q\cdot k}\right ) \right. \nonumber \\
&& \left. +Q_e \frac{\not
\epsilon \not k }{2 p_e\cdot k}\right ]\left (2\not q +m_e\right)(1-\gamma_5)v(p_\nu,
\lambda_\nu)
\label{ke3neutralnewC}. \ea
Then the resulting gauge invariant formula (\ref{ke3neutralnewC}) is not anymore consistent with the leading
logarithm approximation for a photon emission collinear with the $\pi^\pm$.
On the other hand, for the
$K^0$ decay there is no collinear enhancement of photon emission
along the  $\pi^\pm$ direction  because it is not ultrarelativistic.
This is why the inconsistency has no practical consequences for $K^0$ decay but
 may be of importance for the case of
$B_{l3}$ decays.
The case would be then  quite similar
to the one of
 Ref. \cite{Nanava:2009vg} (formulas (11) and (13) there) and is also rather simple to fix
without return to the complete formula (\ref{ke3neutralnew}).
Improved in that respect
formula (\ref{ke3neutralnewC}) would read:
 \ba M_{I^{\prime \prime}}^0 &=& \frac{G_F V_{us}
F_{K\pi}(t)}{\sqrt 2} \overline {u}(p_e, \lambda_e)\Biggl[\left
(Q_e \frac{p_e \cdot \epsilon}{p_e\cdot k} + Q_\pi\frac{q \cdot
\epsilon}{q\cdot k}\right ) \; \nonumber \\
&&
\left (2\Bigl(\not q +\not k \frac{p_e\cdot k}{q\cdot k+ p_e\cdot k}\Bigr) +m_e\right)
 \nonumber \\
&&
 +Q_e \frac{\not
\epsilon \not k }{2 p_e\cdot k}\left (2\not q +m_e\right) \;\Biggr]
 (1-\gamma_5)v(p_\nu,
\lambda_\nu) \label{ke3neutralnewD}. \ea
We   investigate it, as another option  for bremsstrahlung
matrix element in neutral $K_{l3}$ decays.

Our formula ({\ref{ke3neutralnew}) can be re-written  also as a sum
of $M_{I^{\prime \prime}}^0$ and  $M_{{II}^\prime}^0$: \ba
M_0 &=&M_{I^{\prime \prime}}^0 + M_{{II}^\prime}^0 \, , \nonumber \\
M_{{II}^\prime}^0 &=&2\frac{G_F V_{us} F_{K\pi}(t)}{\sqrt 2} \left (
\frac{k^\mu}{q\cdot k+ p_e\cdot k} \left ( Q_\pi q \cdot
\epsilon + Q_e p_e\cdot \epsilon \right )\right. \nonumber \\
&&\left.- Q_\pi\epsilon^\mu \right ) \overline {u}(p_e, \lambda_e)
\gamma_\mu (1-\gamma_5)v(p_\nu, \lambda_\nu)\label{ke30II}\, , \ea
where $M_{{II}^\prime}$ is free of singularities.
\section{Differential decay probability.}
As the spin states of the $K_{l3}$ decay products
are not measurable, we  concentrate on  differential distributions obtained
from amplitudes squared and summing over lepton spin states.
\subsection{Born and real emissions}
Let us  explore squares  of amplitudes given by
eqs.~(\ref{ke3ctotalnewI}, \ref{ke3ctotalnewII}).
 We use the following notations:
\ba S= 2p_e\cdot p_\nu\, , T = 2q\cdot p_e\, ,
U = 2q\cdot p_\nu\, .  \ea
The Born level expression for the charged $K$ decay reads:
\ba \sum_{\rm Spin}|M^c_{\rm Born}|^2 &=&
\frac{G_F^2 |V_{us}|^2 F_{K\pi}^2(t)}{4}32\nonumber \\
&& \left [q\cdot
p_\nu \left (2q\cdot p_e +m_e^2\right)-
\left (m_\pi^2-\frac{m_e^2}{4}\right )p_\nu\cdot p_e\right ] \nonumber \\
&=&\frac{G_F^2 |V_{us}|^2 F_{K\pi}^2(t)}{4}16\nonumber \\ && \left
[U\left (T+m_e^2\right) - \left (m_\pi^2-\frac{m_e^2}{4}\right
)S\right ]\, .
\label{amp_born} \ea
For the bremsstrahlung case, the square of the
amplitude is given by
\ba
 \sum_{\rm Spin}|M^c|^2 &=& \sum_{\rm Spin}|M^c_{I^\prime}|^2 +  \sum_{\rm Spin}|M^c_{II^\prime}|^2 \nonumber \\
  && + 2\sum_{\rm Spin} M^c_{I^\prime}{M^c_{II^\prime}}^* \, ,
\label{squared_chargedK}
\ea
where
\ba
 \sum_{\rm Spin}|M^c_{I^\prime}|^2 &=&32\sum_{i=1, 2}\left (Q_e\frac{p_e\cdot \epsilon_i }{p_e\cdot k} -Q_K
 \frac{p\cdot\epsilon_i}{p\cdot k} \right )^2 \nonumber \\
 && \frac{G_F^2
|V_{us}|^2 F_{K\pi}^2(t)}{4}\left [q\cdot p_\nu\left (
2q\cdot p_e + m_e^2\right) \right. \nonumber \\
  &&\left. - \left (m_\pi^2-\frac{m_e^2}{4}\right )p_\nu\cdot p_e\right ]\, ,
\label{squared_chargedKI}
\ea
\ba
\sum_{\rm Spin}|M^c_{II^\prime}|^2 &=&\frac{-16Q_e^2}{p_e\cdot k}\sum_{i=1, 2}\left (\epsilon_i\cdot \epsilon_i\right )
\frac{G_F^2|V_{us}|^2 F_{K\pi}^2(t)}{4}\nonumber \\
  && \left [2 q\cdot p_\nu q\cdot k - \left
(m_\pi^2-\frac{m_e^2}{4}\right )p_\nu \cdot k \right ]\nonumber \\\label{squared_chargedKII}
\ea
\ba
 2\sum_{\rm Spin} M^c_{I^\prime}{M^c_{II^\prime}}^* && = 32\frac{G_F^2
|V_{us}|^2 F_{K\pi}^2(t)}{4}\, \nonumber \\
&&
\left [\sum_{i=1, 2}Q_e\frac{p_e\cdot \epsilon_i }{p_e\cdot k}
\left (Q_e\frac{p_e\cdot \epsilon_i }{p_e\cdot k} -Q_K
 \frac{p\cdot\epsilon_i}{p\cdot k} \right )\right. \nonumber \\
 && \left.
\left (2q\cdot p_\nu q\cdot k -
 \left (m_\pi^2-\frac{m_e^2}{4}\right ) p_\nu \cdot k \right
 )\right. \, \nonumber \\
 && \left.  -\sum_{i=1, 2} Q_e
\left (Q_e\frac{p_e\cdot \epsilon_i }{p_e\cdot k} -Q_K
 \frac{p\cdot\epsilon_i}{p\cdot k} \right ) \right. \nonumber \\
 && \left. \left ( 2q\cdot p_\nu q\cdot \epsilon_i  -
\left (m_\pi^2-\frac{m_e^2}{4}\right ) p_\nu \cdot \epsilon_i\right)
\right]\, . \nonumber \\\label{squared_chargedKIII}
 \ea
Here $\epsilon_1$, $\epsilon_2$ are two orthogonal photon polarization vectors.
As  expected $\sum|M^c_{I^\prime}|^2$ consists of a
Born-like expression  multiplied by an eikonal factor. The second and third terms,
$\sum|M^c_{II^\prime}|^2$ and $2\sum
M^c_{I^\prime}{M^c_{II^\prime}}^*$, are free of soft singularities.

Formulae for charged $K^-$ decay and neutral  $K^0\to\pi^ + e^- \bar\nu_e$ are similar.
Obtained from formula~(\ref{born-neutral})
\ba \sum_{\rm Spin}|M^0_{\rm Born}|^2 = 2\times \sum_{\rm Spin}|M^c_{\rm Born}|^2\,  \ea
differs from the Born contribution of charged  $K$  decay by a factor of 2.
For the bremsstrahlung case, the  amplitude squared
can be again separated into parts. The first one consists\footnote{First term can be  obtained  from formula
(\ref{squared_chargedKI},\ref{squared_chargedKII},\ref{squared_chargedKIII})
with the help of the change of variables
\ba Q_K\to - Q_\pi\, , \ \ \
\frac{p\cdot\epsilon_i}{p\cdot k}\to \frac{q\cdot\epsilon_i}{q\cdot
k}\, . \ea
} of the squared formula
(\ref{ke3neutralnewD}), which reads:
 \ba
 &&\sum_{\rm Spin} |M_{I^{\prime \prime}}^0|^2 =2\times {\sum_{\rm Spin} |M^c|^2}_{Q_K \rightarrow - Q_\pi,
 \frac{p\cdot\epsilon_i}{p\cdot k}\rightarrow \frac{q\cdot\epsilon_i}{q\cdot k} }  \nonumber \\
 && + 32\frac{G_F^2
|V_{us}|^2 F_{K\pi}^2(t)}{2}\frac{p_e\cdot k}{q\cdot k+ p_e\cdot
k}\times \nonumber \\
&& \left [\sum_{i=1, 2}\left (Q_e\frac{p_e\cdot \epsilon_i
}{p_e\cdot k} +
 Q_\pi \frac{q\cdot\epsilon_i}{q\cdot k} \right )^2 \left ( k\cdot p_\nu \left (2 q\cdot p_e +m_e^2\right )\right. \right.\nonumber \\
 && \left.\left. + 2\left( q\cdot p_\nu
 + \frac{p_e\cdot k}{q\cdot k+ p_e\cdot k}k\cdot p_\nu\right ) k\cdot
p_e - 2q\cdot k p_\nu\cdot p_e\right )\right. \nonumber \\
&& \left. + \sum_{i=1, 2} 2 Q_e \left (Q_e\frac{p_e\cdot \epsilon_i}
{p_e\cdot k} + Q_\pi \frac{q\cdot\epsilon_i}{q\cdot k} \right )\right. \nonumber \\
 && \left. \left (p_\nu \cdot \epsilon_i q\cdot k - q\cdot \epsilon_i p_\nu
\cdot k \right )\right ]\, .
\ea

The second part, contributing to  the squared amplitude for neutral $K$ decay,  consists of the
squared  formula (\ref{ke30II}) and its interference with formula
(\ref{ke3neutralnewD}).
We do not write explicitly this expression here, as it is rather lengthy,
and is numerically  small.
\subsection{Virtual corrections}
Virtual corrections to $K_{l3}$ decays  can be found eg. in Ref.~\cite{Cirigliano:2008wn}, they are of lesser importance
than the real ones and we do not recall them here.
The present work  is devoted to the discussion of  real
emission corrections, which are experimental condition dependent.
Virtual corrections have to be divided into two parts. One part
 is large, but  adds up to zero with real emissions in total rate thanks
to Kinoshita-Lee-Nauenberg theorem. This part is  taken into account by the {\tt PHOTOS}
Monte Carlo code. The other part has to be included  in the form-factor and incorporated to the Born
level matrix
element. For this purpose, real emission amplitude squared need to be integrated
over photon momentum
 to control the sum rule to the level of complete first order.
This can be performed  in an approximate way\footnote{%
But then,
no gain beyond  Konoshita-Lee-Nauenberg can be achieved. In $K_{e3}$ decay there is no Coulomb effect.}  (as in Ref. \cite{Bystritskiy:2009iv}) or in an exact manner, following
phase space parametrization as used in {\tt PHOTOS}.
Integration (analytic or numerical) need to be performed.
This solution has to be adopted, once experimental precision  approaches $\frac{\alpha}{\pi}$
precision level. Still another type of solution, non exploiting the sum rule of   Konoshita-Lee-Nauenberg can be
useful.  We mention such possibility in  footnote \ref{foot:semi}.

\section{Monte Carlo Simulation}

  We use the {\tt TAUOLA} \cite{Jezabek:1991qp} code to generate Born level $K$ decay samples.  For practical reasons
this generator is suitable
for our purpose once the $\tau$ decay matrix
element is replaced with the one of the $K_{l3}$ decay, and appropriate adjustment of masses and particle identifiers are performed.
 Semileptonic decays of $\tau$'s are suitable for such adaptation\footnote{ \label{foot:semi} {\tt TAUOLA} semileptonic
decay channel
offer an  alternative crude phase space generator for radiative corrections,
which may become useful in the future, especially if complete virtual
corrections are
to be included and studied.}.
Another advantage is that we can then guarantee full control of parameters
used in our Born level generator and in  matrix element of {\tt PHOTOS} correcting weights.

We gain, because  {\tt PHOTOS} is ready to use
with {\tt TAUOLA}. The two programs share  technical elements. This is convenient when non-factorizable
parts of matrix elements are installed.
The
interface to {\tt HEPEVT} event record of the two programs offers an easy access to our testing
tool {\tt MC-TESTER} \cite{Golonka:2002rz,Davidson:2008ma}.
The prepared plots  are
 then compatible with our previous studies. From the user point of view, numerical results
collected for this paper, and available also in graphic form from the web-page \cite{web-Kl3},
 can be of interest for benchmarking  $K$ decay
generators independently, whether  they are coded in {\tt FORTRAN} or in C++.

\begin{figure}[htp!]
\begin{tabular}{ccc}
\subfigure[$M_{\pi^0e^-}$]{
\includegraphics[%
  width=0.40\columnwidth]{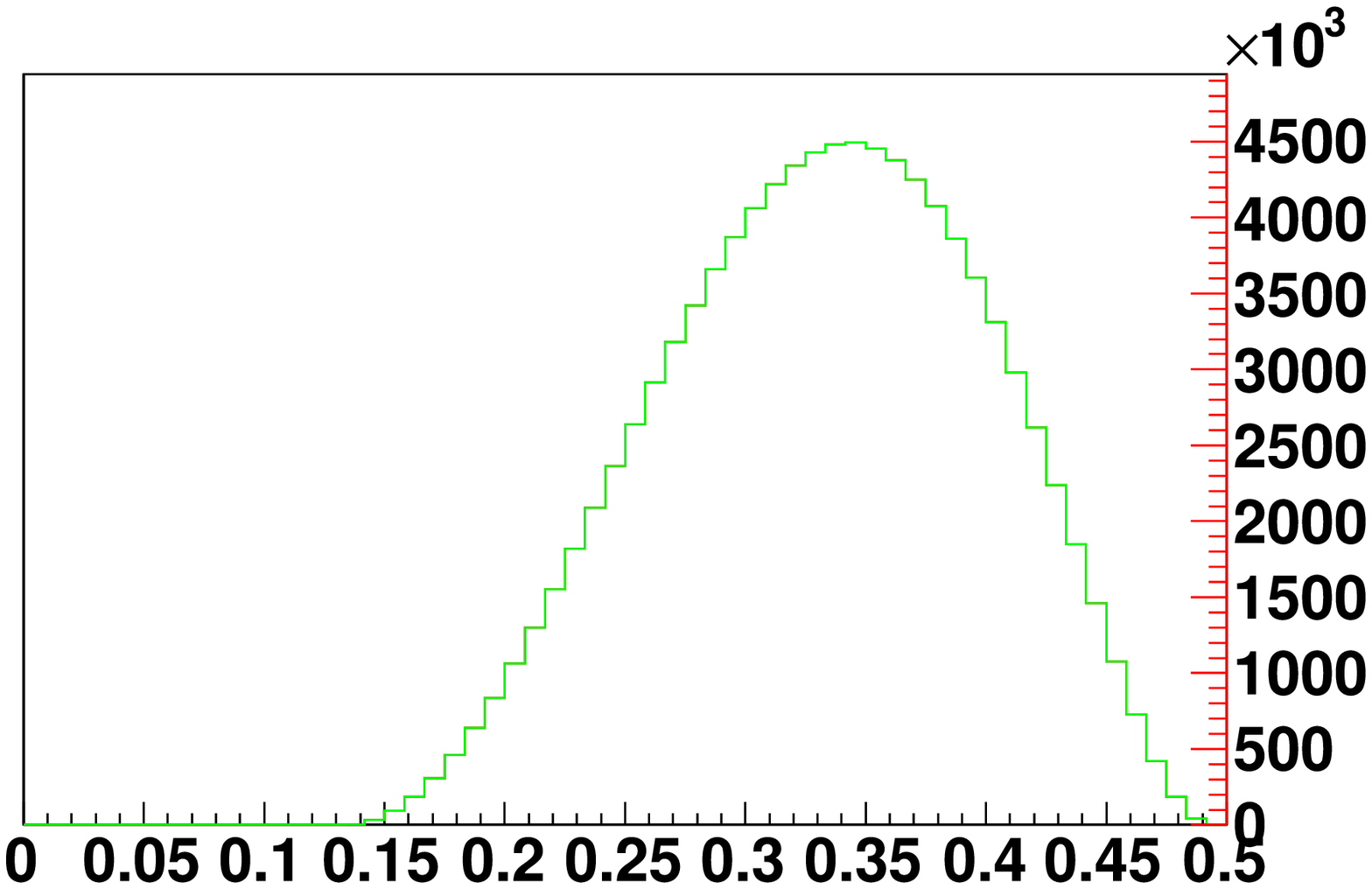}} & \subfigure[$M_{\pi^0\bar\nu_e}$]{\includegraphics[%
  width=0.40\columnwidth]{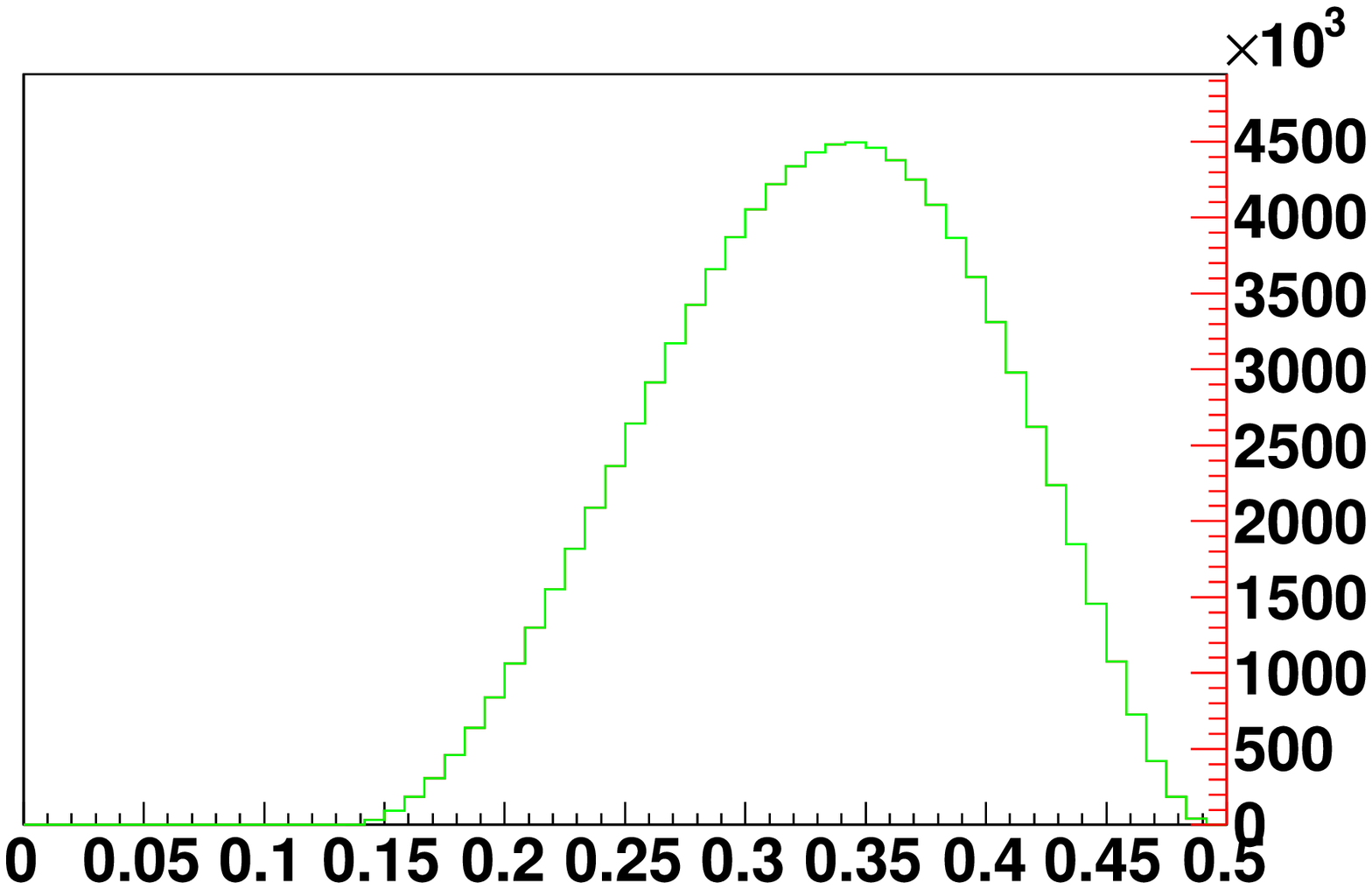}} \\
\subfigure[$M_{e^-\bar\nu_e}$]{  \includegraphics[%
  width=0.40\columnwidth]{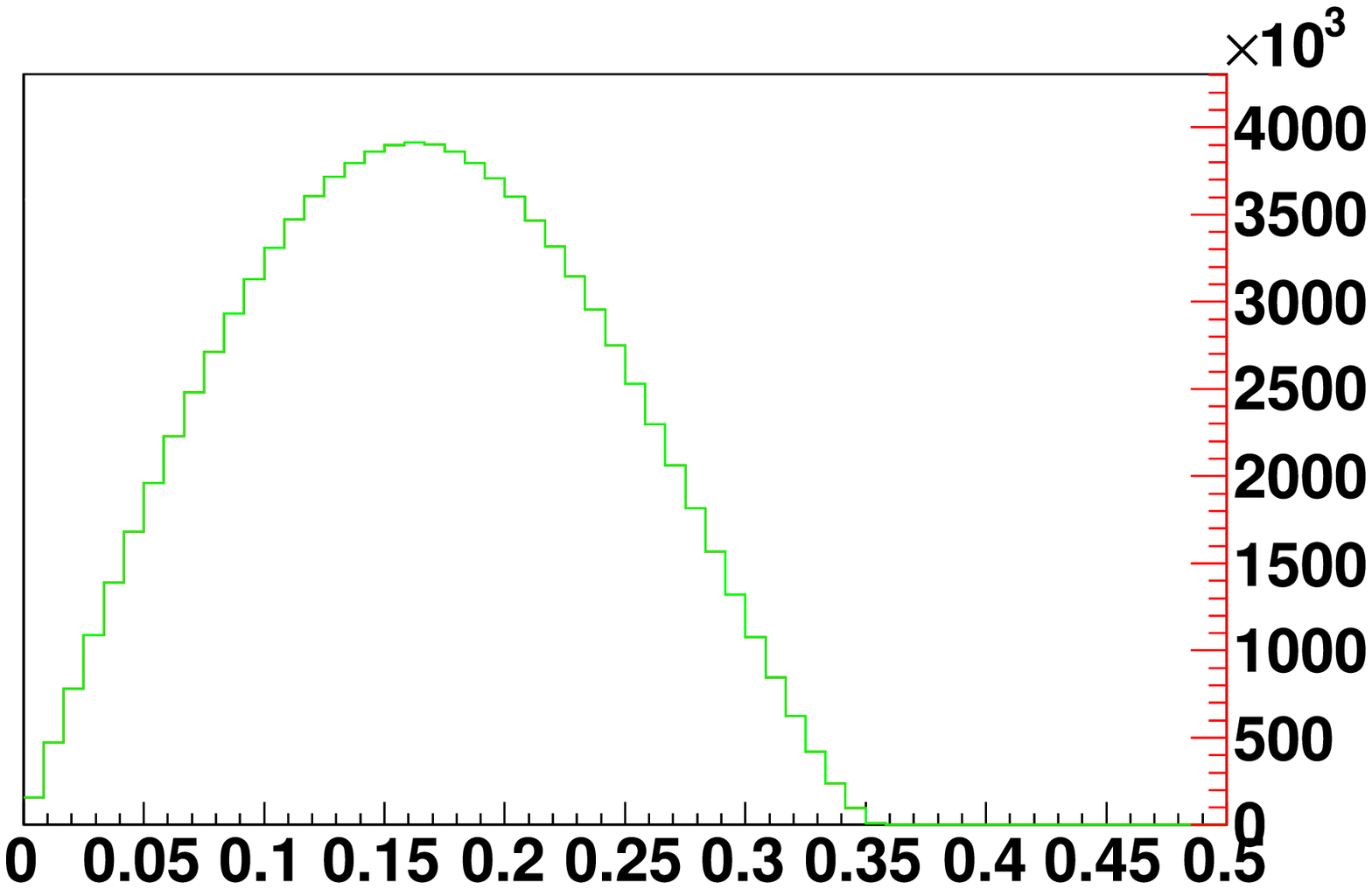}}
\end{tabular}
\caption{Distributions of invariant masses, in GeV (GeV/$c^2$, $c=1$) units, constructed from the  products of the decay $K^-\to \pi^0 e^- \bar\nu_e$,
at Born level.  As in~\cite{Nakamura:2010zzi} $V_{us}=0.2252$ is used, we take however
 $F_{K\pi}(t) = 1$. This is an  acceptable approximation for our purposes
and consistent with Ref.\cite{Cirigliano:2008wn}. \label{figborn}}
\end{figure}

The squares of Born level amplitudes (\ref{amp_born}) for  decays $K^-\to \pi^0 e^- \bar\nu_e$,
and $K^0\to \pi^+ e^- \bar\nu_e$ are rather easy to implement into {\tt TAUOLA}.
The numerical results, as histograms of invariant masses constructed from pairs of final state decay products,
 are shown in Fig.~\ref{figborn}. We have compared these
results with the one of Ref.~\cite{Batley:2006cj} and reasonable agreement
was found. This comparison is
sufficient for tests, because the matrix element is rather simple and {\tt TAUOLA} itself is well tested.
 Note that detector acceptance effects were included
in Ref.~\cite{Batley:2006cj}, therefore the corresponding figures  do not
coincide
in all details with our  Fig.~\ref{figborn}.

\section{Numerical results}

In all tests presented in  this paper we use samples of 100 milion events.
We refer
 to standard {\tt PHOTOS} whenever we  use
its publicly available {\tt FORTRAN} version 2.15, or any other version
which yields  equivalent results.
In particular, identical results are available (as default option)
from the C++  {\tt PHOTOS} \cite{Davidson:2010ew},
 version 3.0 or higher. One of the  goals of the present work is
to provide a systematic error for these widely used versions.

As a first step, we  perform a
technical test.
For the decay of charged $K$  we have compared results of the standard {\tt PHOTOS} with the {\tt PHOTOS} version of \cite{Nanava:2006vv}.
It was the first version where
the multiphoton phase space generation for final state
of a single charged (and scalar) particle was exact, which was provided with the help
of explicit  phase space Jacobians. It is  publicly available
starting from {\tt PHOTOS++}, version 3.3.
We could  see that
the numerical effect is small. As expected, the difference
is below 0.05 \% if calculated with respect to the  total rate.
We obtained
differences at
the level of 10 \% in regions of phase space contributing at the level of
 $10^{-4}$ to the total rate.

In a second step,
we have checked  the contribution from  collinear photon emission region.
To this end we have selected only those events, used in the previous comparison, where the
photon-electron pair invariant mass was
at most 0.01  of their  energies product (taken in the rest frame of  $K$).
We obtained perfect agreement, with
no statistically significant differences.

Only then, results presented in our article were prepared.
Throughout  the paper, we use our testing program
{\tt MC-TESTER} \cite{Golonka:2002rz,Davidson:2008ma}. The two colored (grey) lines correspond
to the  compared generation samples\footnote{ The distributions of Lorentz invariants constructed from outgoing
particles are shown.  Additional information is available on plots
of  the web-page \cite{web-Kl3}.
The SDP (shape difference parameter) obtained from {\tt MC-TESTER} represents an exclusive
surface (normalized to unity)
under the green and red distributions. The
statistical error (calculated in a rather simplified way) is subtracted from this difference.
The black line is the ratio of the distributions.  }.
To define the boundary of the real emission phase space we
use the photon energy in the rest frame of the decaying kaon: it is set at 0.005 of the decaying kaon mass.

We have repeated the same comparison as discussed previously
but when the complete scalar QED matrix element is
installed. As we can see  from Fig.  \ref{figME} the numerical effect
of the Matrix Element and exact phase space implementation is rather small,
visible only at
the ends of the spectra (contributing  at the level of $10^{-3} $ to the total rate) where relative differences are sizable. There,
 matrix element effects are rather substantial and should be kept in mind in
some contexts, e.g. for generating background to other decays.

\begin{figure}[htp!]
\begin{tabular}{ccc}
\subfigure[$M_{\pi^0\gamma}^2$]{\includegraphics[%
  width=0.48\columnwidth]{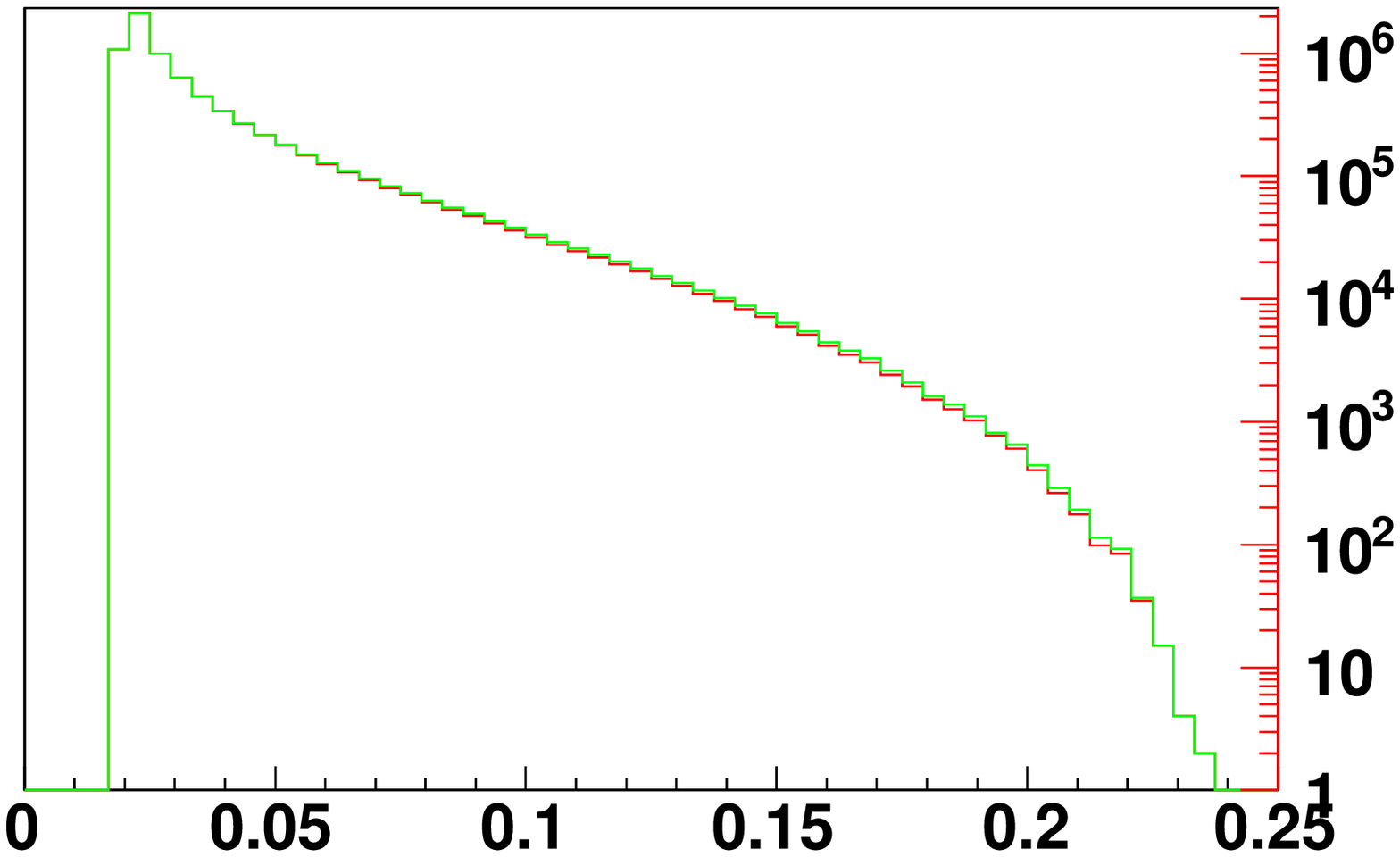}} &
\subfigure[$M_{e^-\gamma}^2$]{\includegraphics[%
  width=0.48\columnwidth]{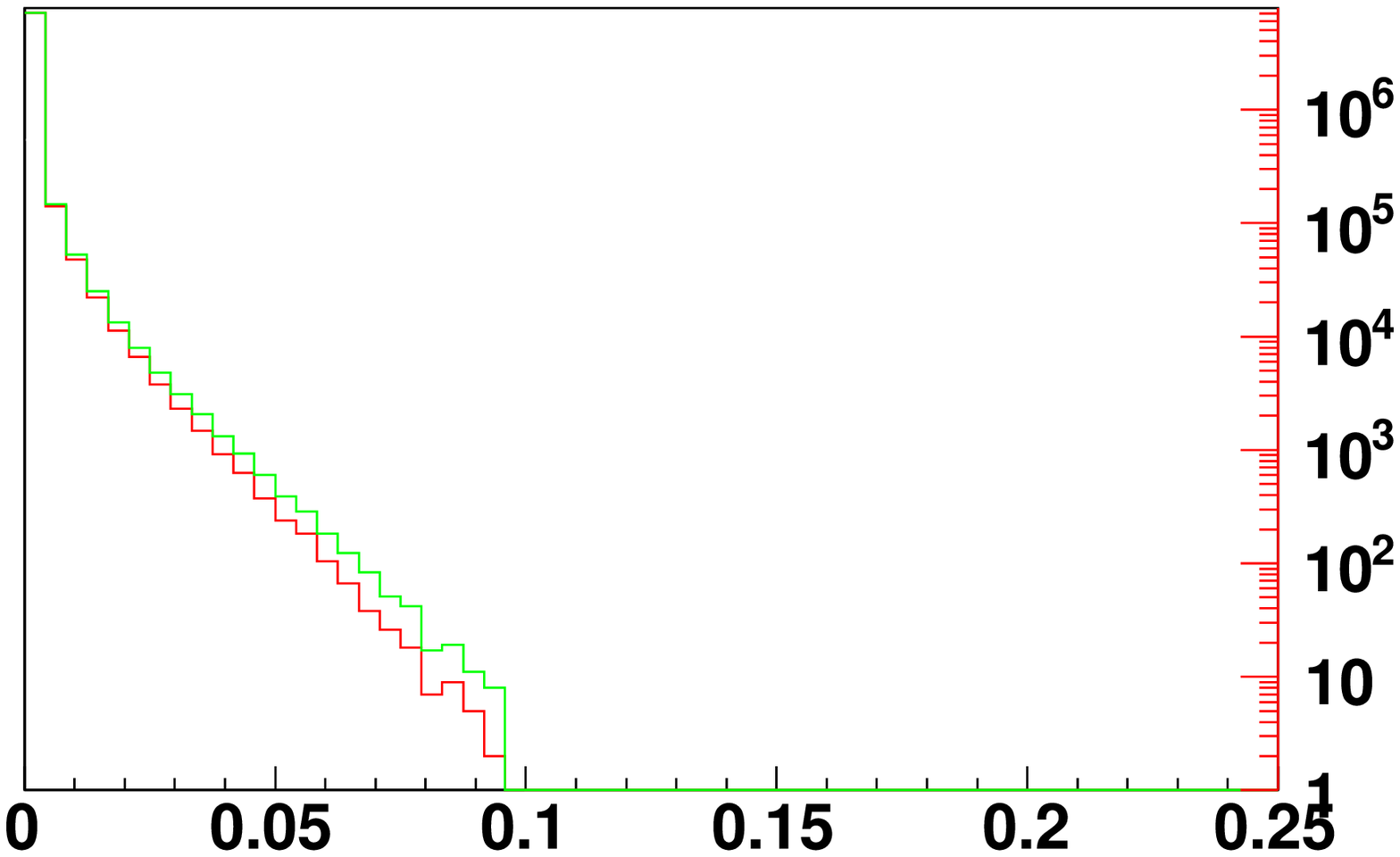}}
\\
\subfigure[$M_{\bar\nu_e\gamma}^2$]{\includegraphics[%
 width=0.48\columnwidth]{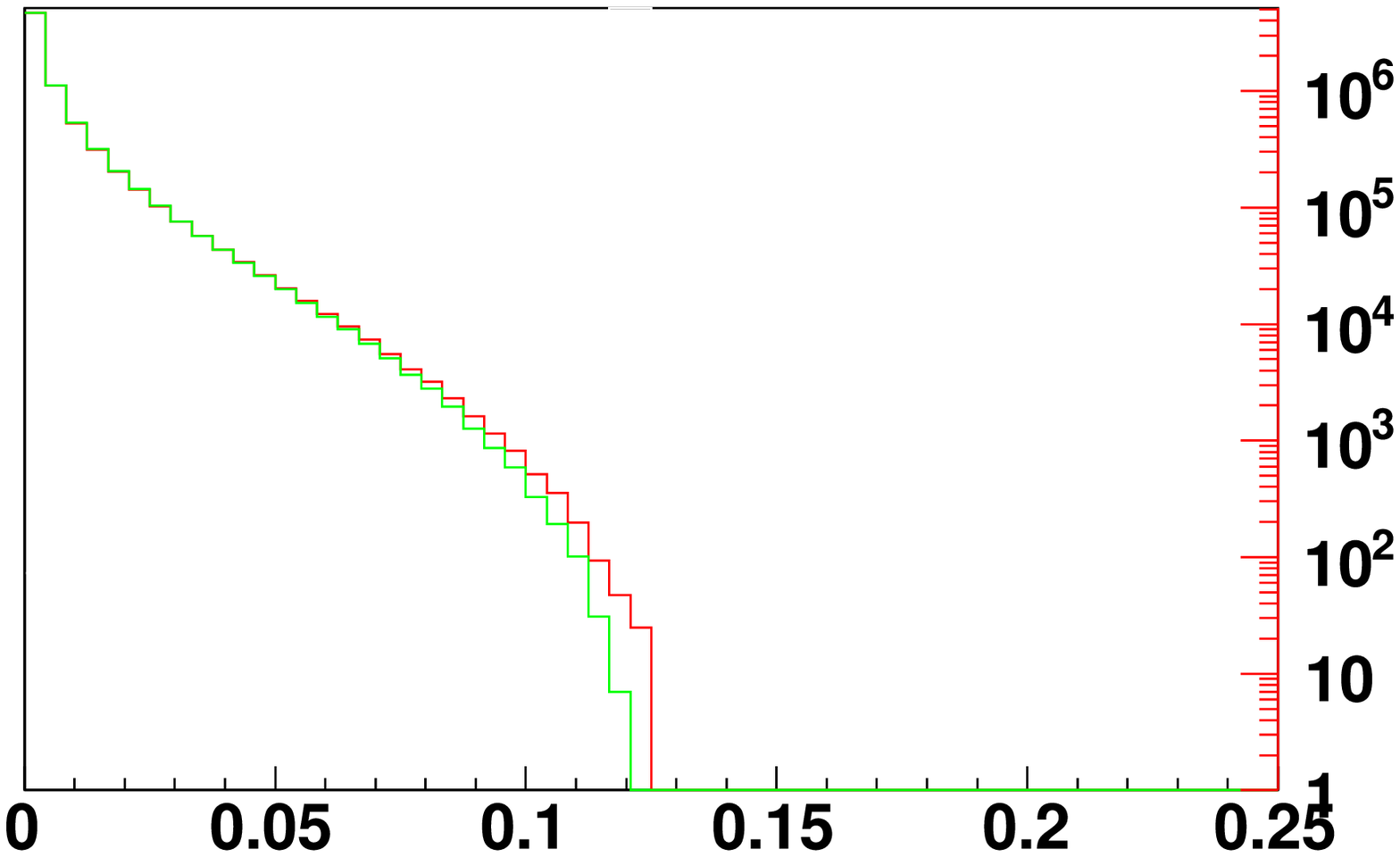}} &
\subfigure[$M_{e^-\bar\nu_e\pi^0}^2$]{\includegraphics[%
   width=0.48\columnwidth]{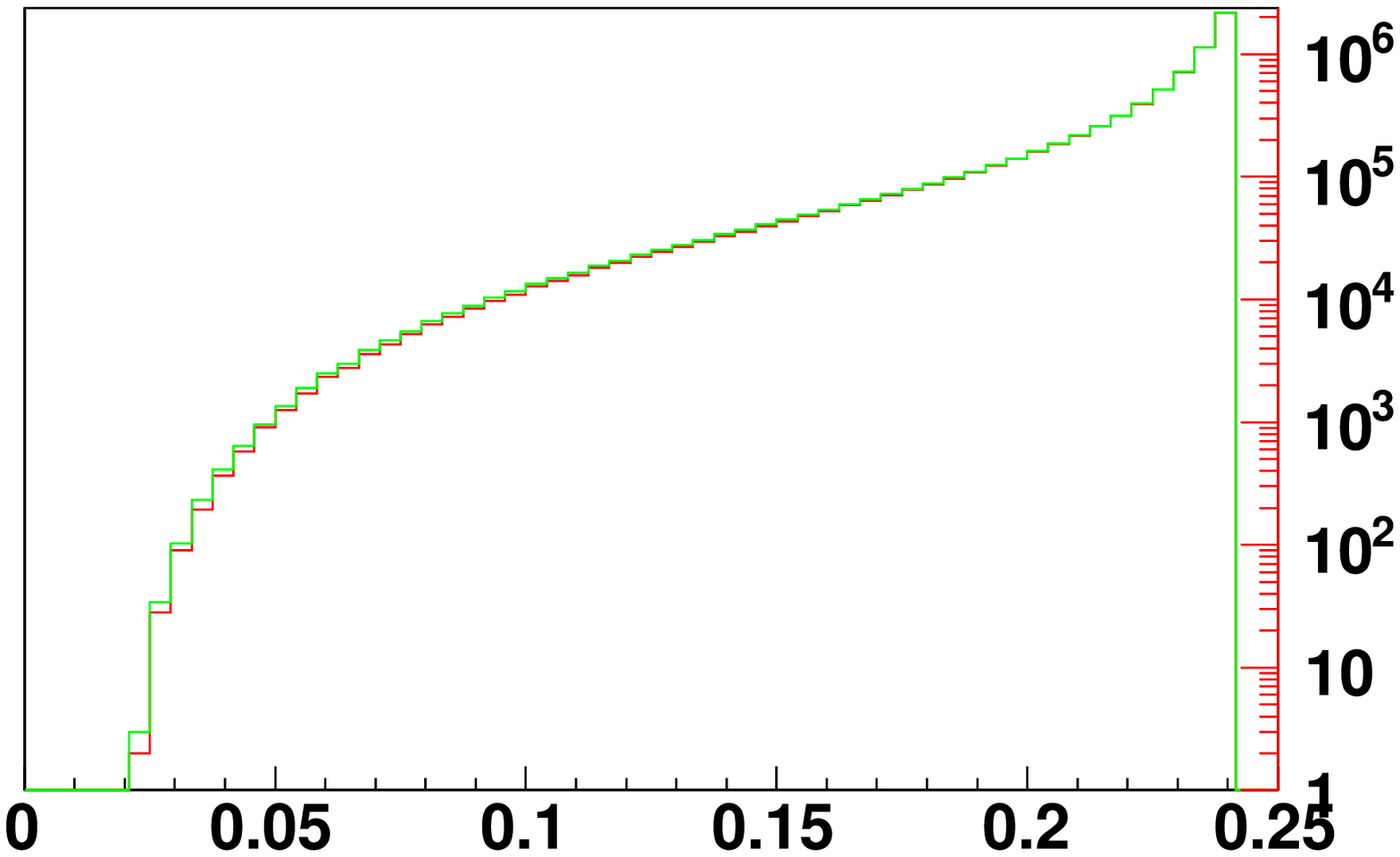}}
\end{tabular}
\caption{Distributions of scalar Lorentz invariants, in GeV$^2$
 (GeV$^2/c^4$, $c=1$) units,  constructed from the decay
products in $K^-\to \pi^0 e^- \bar\nu_e$ channel.
 The most  sensitive invariants to photon energy are plotted. The
red (darker grey) line is standard {\tt PHOTOS}, the green is with exact
Matrix Element.
 The fraction of accepted bremsstrahlung events is
(7.371 $\pm$0.0027) \%
 in standard {\tt PHOTOS} run and
(7.4127 $\pm$ 0.0027) \%
when the matrix element is used.
\label{figME}}
\end{figure}

For the case of $K^-\to \pi^0 \mu^- \bar\nu_\mu$,
logarithmic corrections are of course much smaller. That is why  non leading terms contribute
to photon spectra
in a relatively larger manner, and seemingly
much larger effects can be seen in Fig.~\ref{figMU} than in Fig.~\ref{figME}.
Nonetheless it is not more significant
numerically.
Only 0.45 \% of events enter the plots for the muonic channel;
twenty times less than in the electron case.
 We can conclude that standard {\tt PHOTOS} works sufficiently well for the
 $K^-\to \pi^0 \mu^- \bar\nu_\mu$ decay, if one is interested in 0.2 \% precision limits.

The $K^0\to \pi^+ l^- \bar\nu_l$ case
is technically more interesting as two kinematical branches are present in the crude level
phase space generator.

In Fig.~\ref{figK0e} we compare standard {\tt PHOTOS} with a version where
the scalar QED matrix element is installed.
The approximation in the phase space  is still present. Only
in Fig.~\ref{figK0e1} we use   single
channel presampler for the  phase space generation and the phase space is exact.
The
effect of phase space Jacobian approximation is rather small.

By comparing Figs. \ref{figK0e2} and \ref{figK0e1}
we can see that
 the standard {\tt PHOTOS} is much
closer to the result of scalar QED matrix element than to one of (\ref{ke3neutralnewC}). 
The bulk of the difference is due to the non-compatibility of formula (\ref{ke3neutralnewC})
with collinear logarithms
due to emission from charged pion (we can see it by comparing Fig.~\ref{figK0e3}). This is of course beyond the framework of approximation at use, 
  but it is nonetheless of some interest, to understand the origin
of the residual differences between possible options for the matrix element and how to fix it in gauge invariant way,
but without study of the whole matrix element.
\begin{figure}[htp!]
\begin{tabular}{ccc}
\subfigure[$M_{\pi^0\gamma}^2$]{\includegraphics[%
  width=0.48\columnwidth]{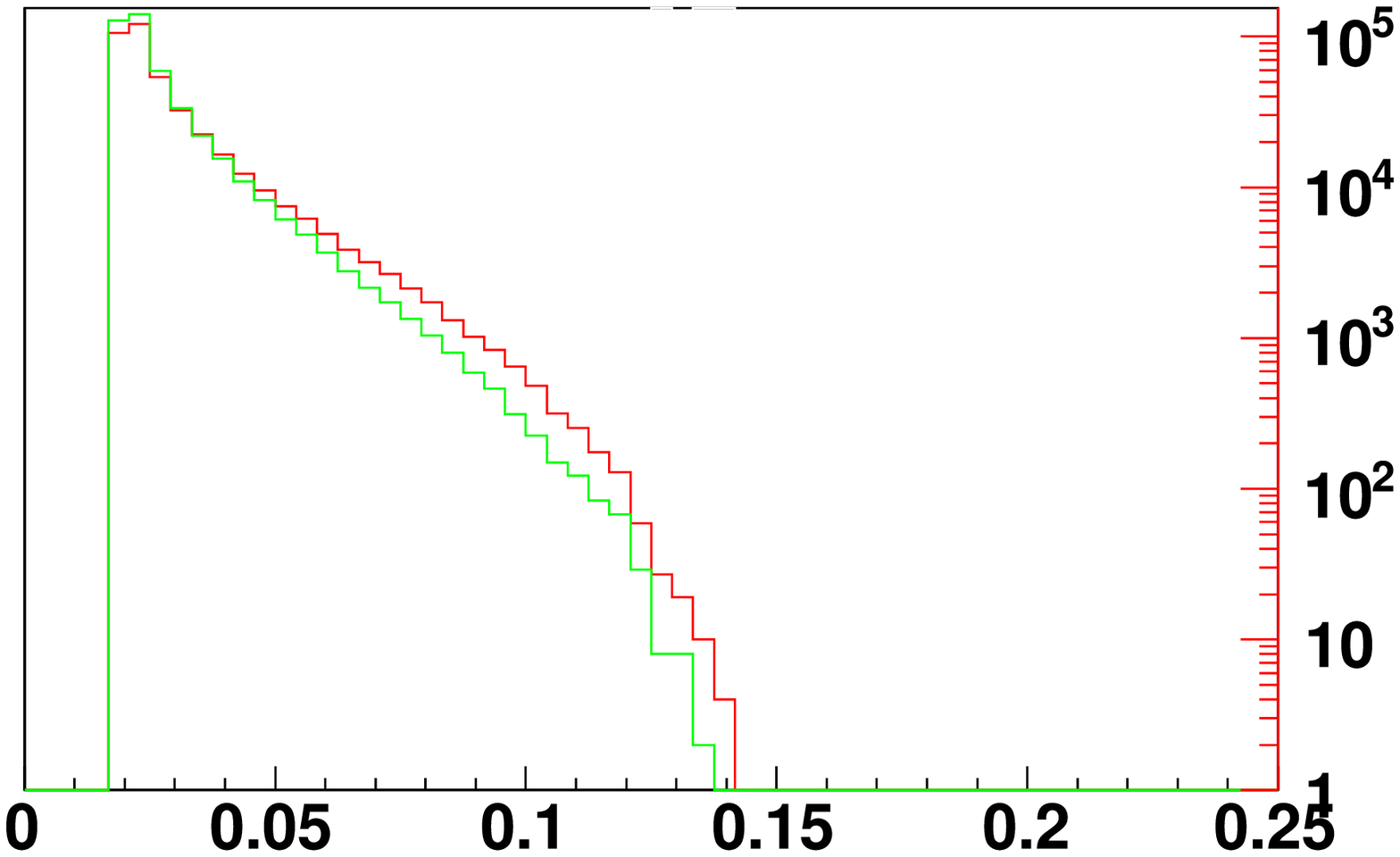}} &
\subfigure[$M_{\mu^-\gamma}^2$]{\includegraphics[%
  width=0.48\columnwidth]{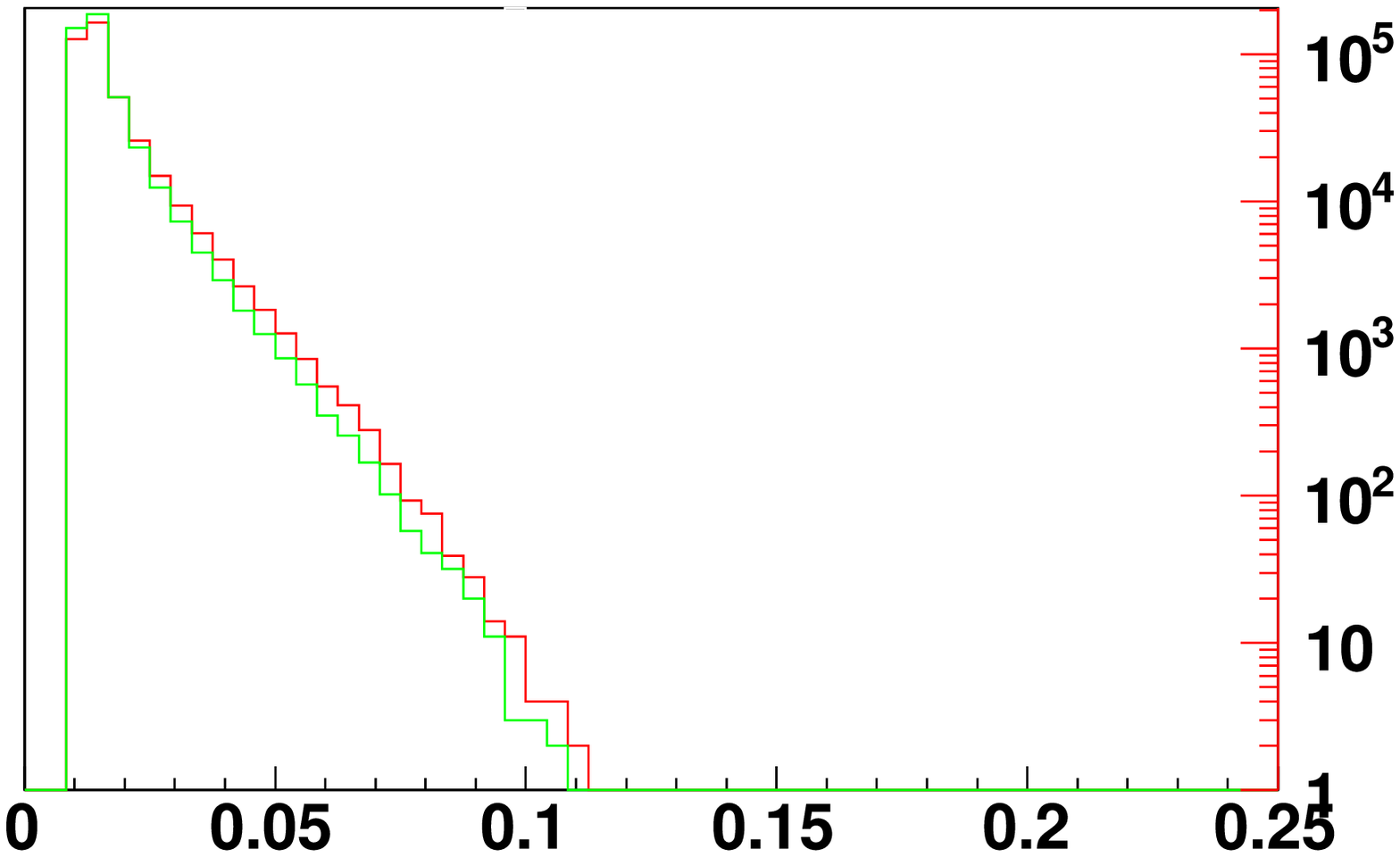}}
\\
\subfigure[$M_{\bar\nu_\mu\gamma}^2$]{\includegraphics[%
 width=0.48\columnwidth]{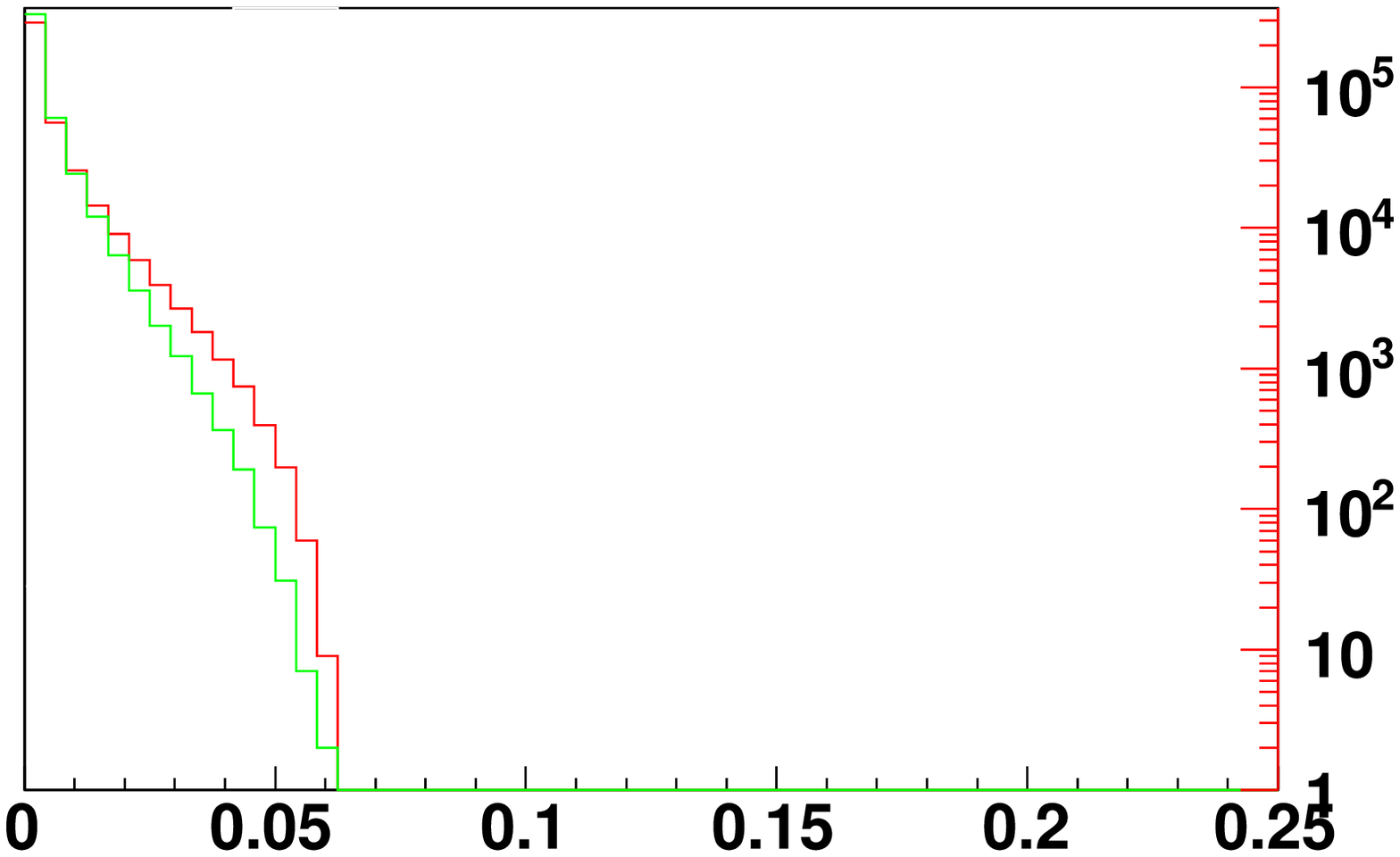}} &
\subfigure[$M_{\mu^-\pi^0\bar\nu_\mu}^2$]{\includegraphics[%
   width=0.48\columnwidth]{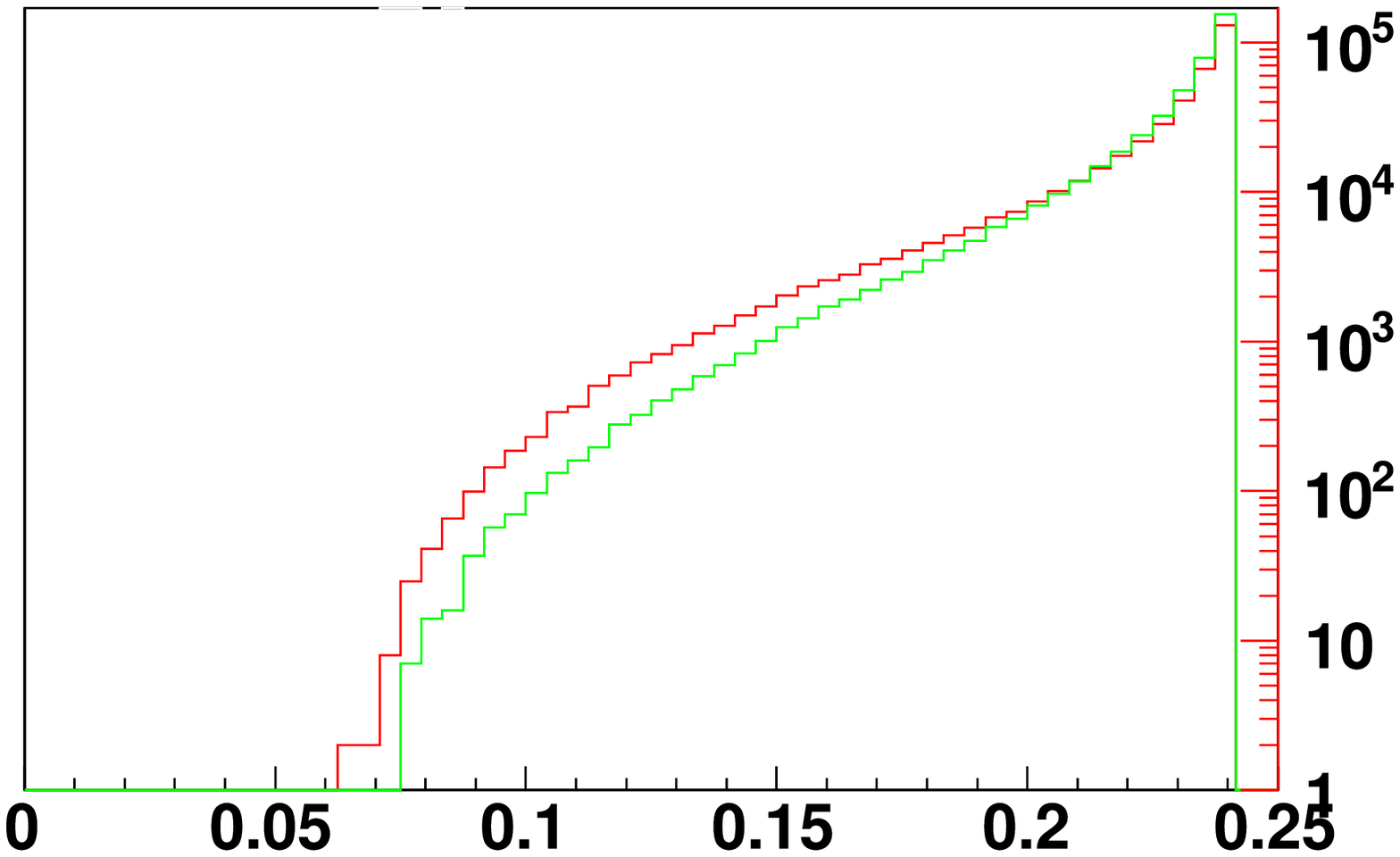}}
\end{tabular}
\caption{Distributions of scalar Lorentz invariants, in GeV$^2$
(GeV$^2/c^4$, $c=1$) units,  constructed from the decay
products in $K^-\to \pi^0 \mu^- \bar\nu_\mu$ channel.
Invariants most  sensitive to photon energy are shown.
The red (darker grey) line is standard  {\tt PHOTOS}, the green is with exact
Matrix Element.
  One could conclude
that the effect of matrix element introduction is not small in this case.
However, only a small fraction of events enter this plot
  (0.4113 $\pm$ 0.0006) \%  for standard {\tt PHOTOS}
and  (0.4445 $\pm$ 0.0007) \% for the  case with matrix element.
The difference is well below 0.1 \% when compared to the total rate.
The two distributions coincide in the soft photon region.
 \label{figMU}}
\end{figure}
\begin{figure}[htp!]
\begin{tabular}{ccc}
\subfigure[$M_{\pi^+\gamma}^2$]{\includegraphics[%
  width=0.48\columnwidth]{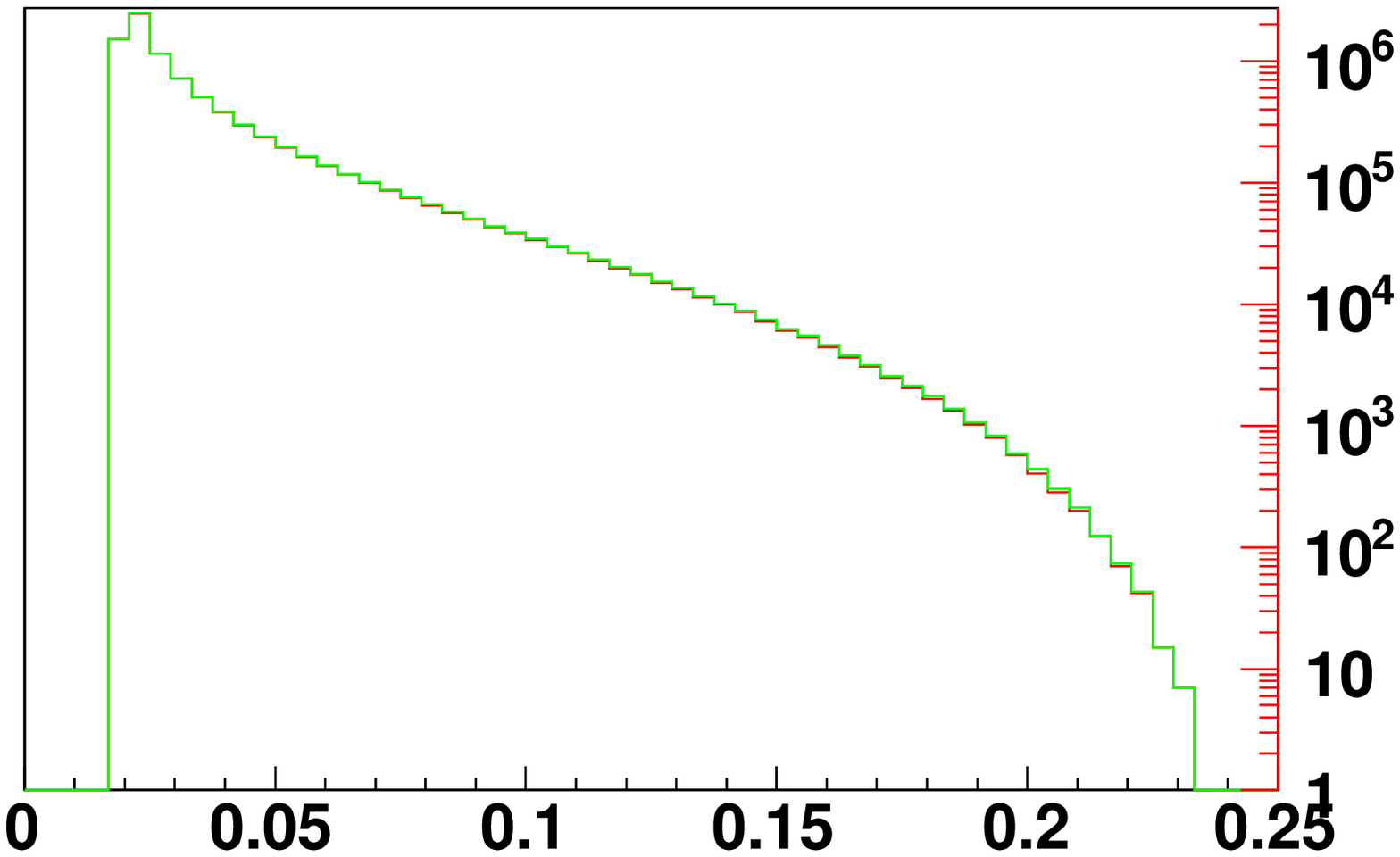}} &
\subfigure[$M_{e^-\gamma}^2$]{\includegraphics[%
  width=0.48\columnwidth]{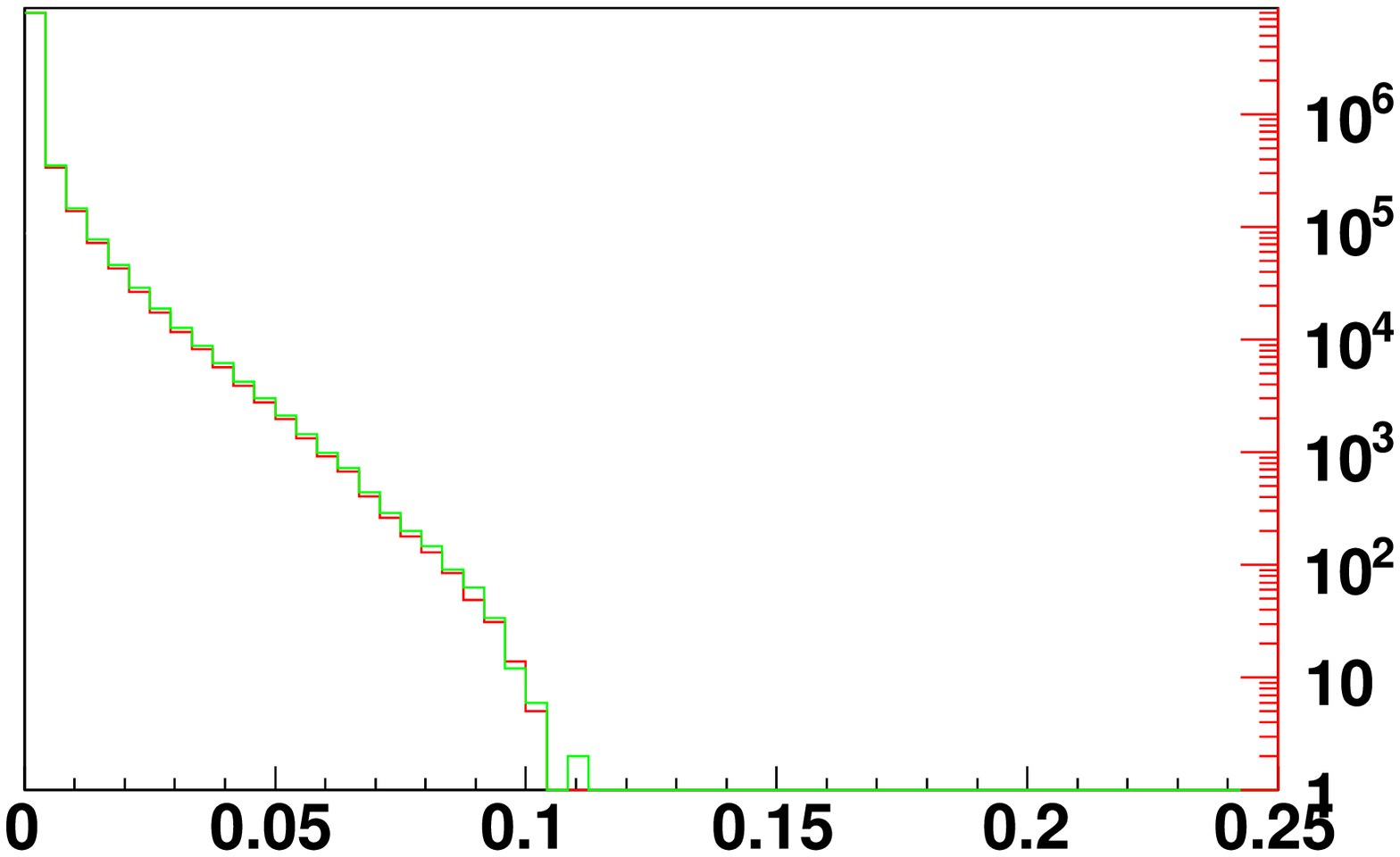}}
\\
\subfigure[$M_{\bar\nu_e\gamma}^2$]{\includegraphics[%
 width=0.48\columnwidth]{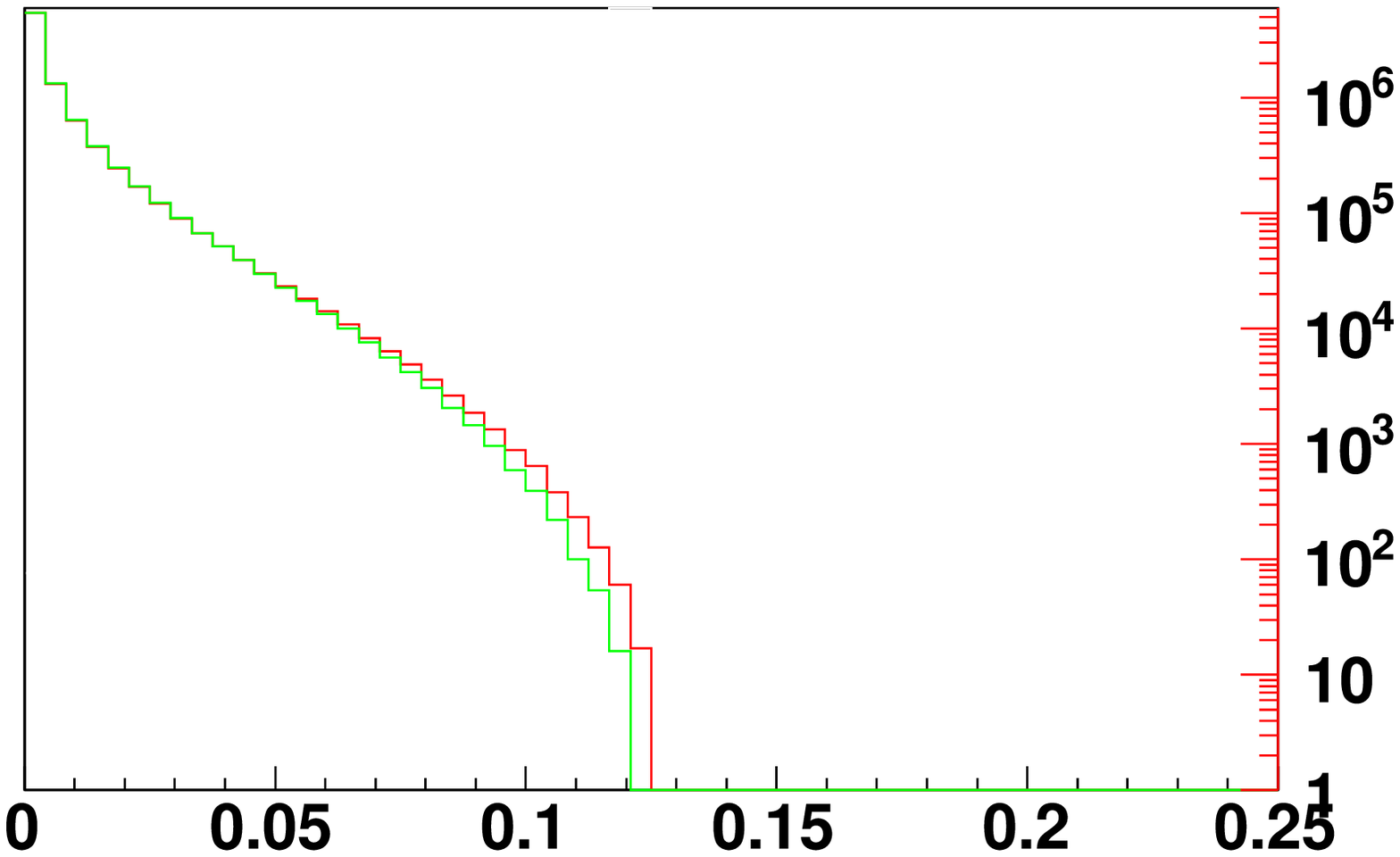}} &
\subfigure[$M_{\pi^+\bar\nu_ee^-}^2$]{\includegraphics[%
   width=0.48\columnwidth]{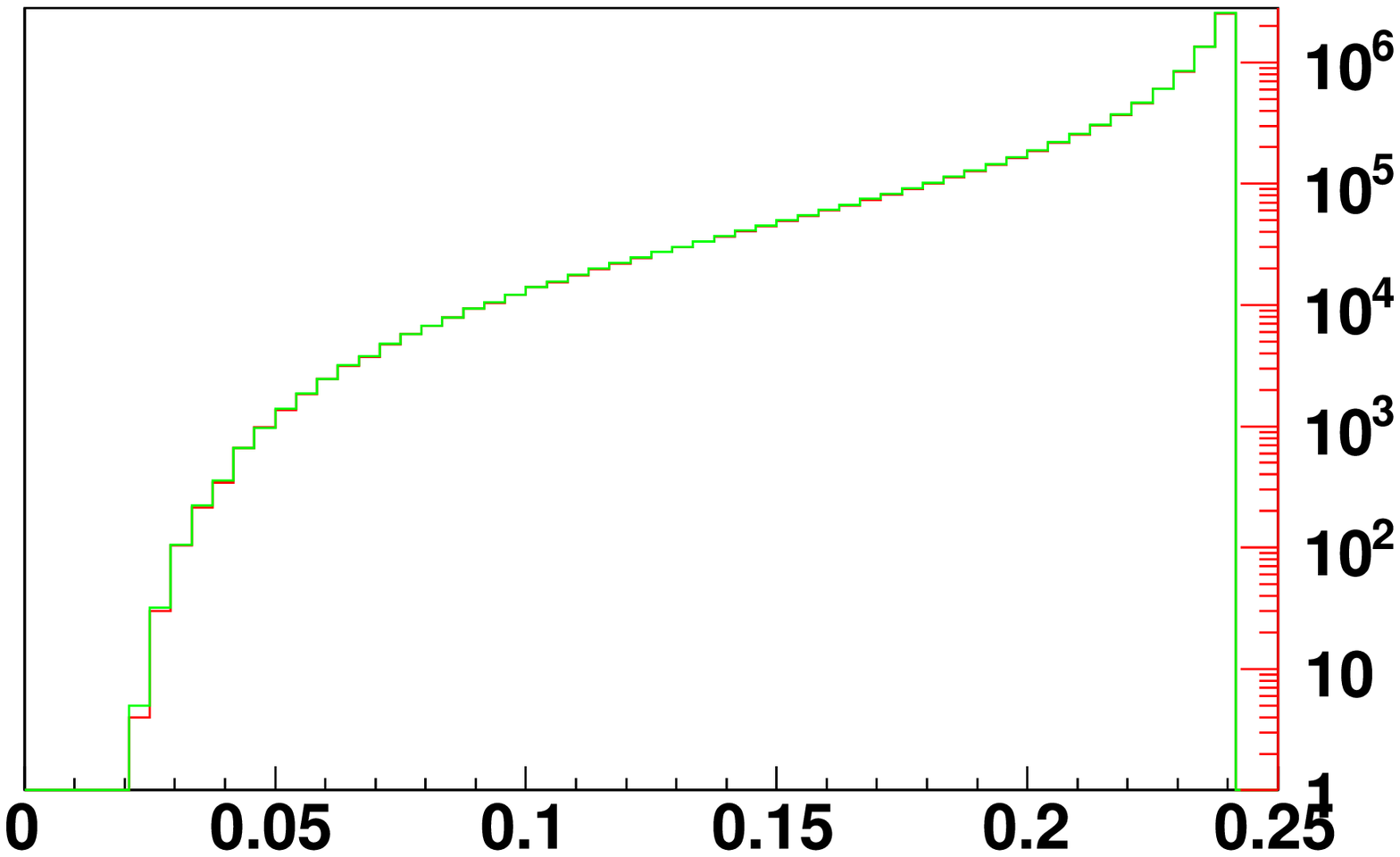}}
\end{tabular}
\caption{Distributions of scalar Lorentz invariants, in GeV$^2$
(GeV$^2/c^4$, $c=1$) units,  constructed from the decay
products in  $K^0\to \pi^+ e^- \bar\nu_e$ channel. The  most  sensitive invariants to photon energy are plotted.
Two kinematical branches are used, thus the phase space is not exact.
The
red (darker grey) line is standard {\tt PHOTOS}, the green is with exact
Matrix Element.
 The fraction of accepted bremsstrahlung events is
 (8.6398 $\pm$ 0.0029) \%
 in standard {\tt PHOTOS} run
and  (8.6913 $\pm$ 0.0029) \% when the matrix element is used. \label{figK0e}}
\end{figure}
\begin{figure}[htp!]
\begin{tabular}{ccc}
\subfigure[$M_{\pi^+\gamma}^2$]{\includegraphics[%
  width=0.48\columnwidth]{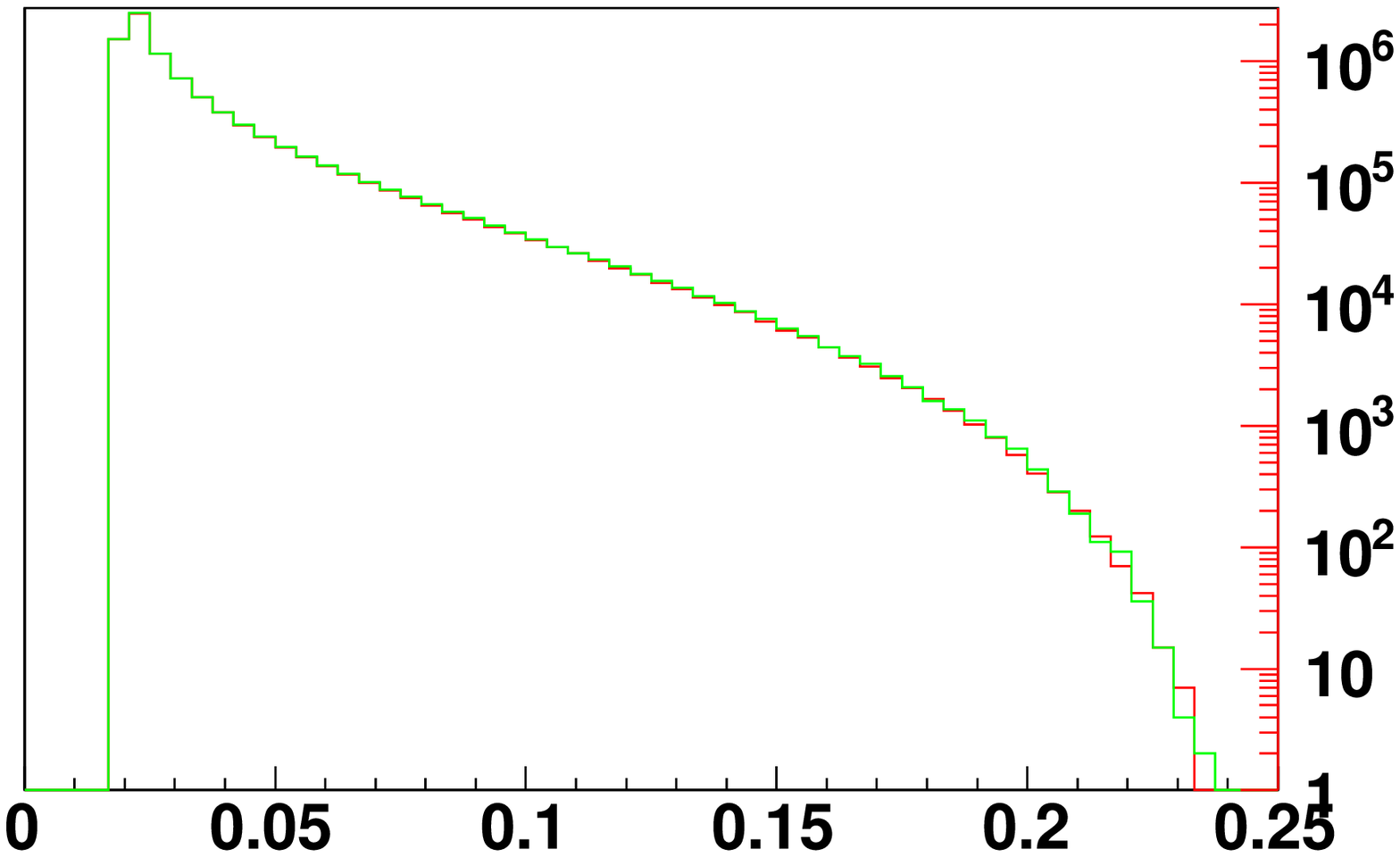}} &
\subfigure[$M_{e^-\gamma}^2$]{\includegraphics[%
  width=0.48\columnwidth]{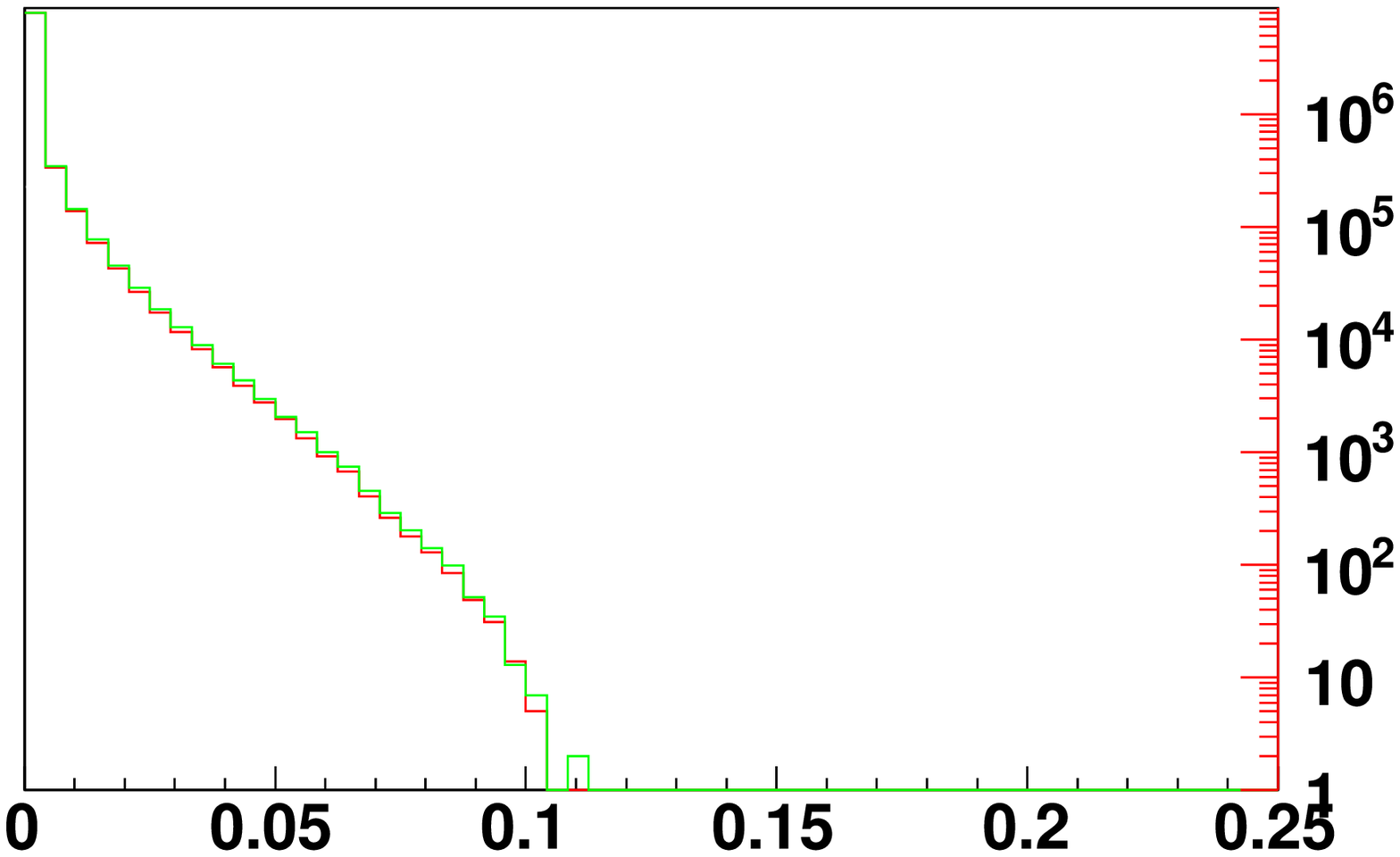}}
\\
\subfigure[$M_{\bar\nu_e\gamma}^2$]{\includegraphics[%
 width=0.48\columnwidth]{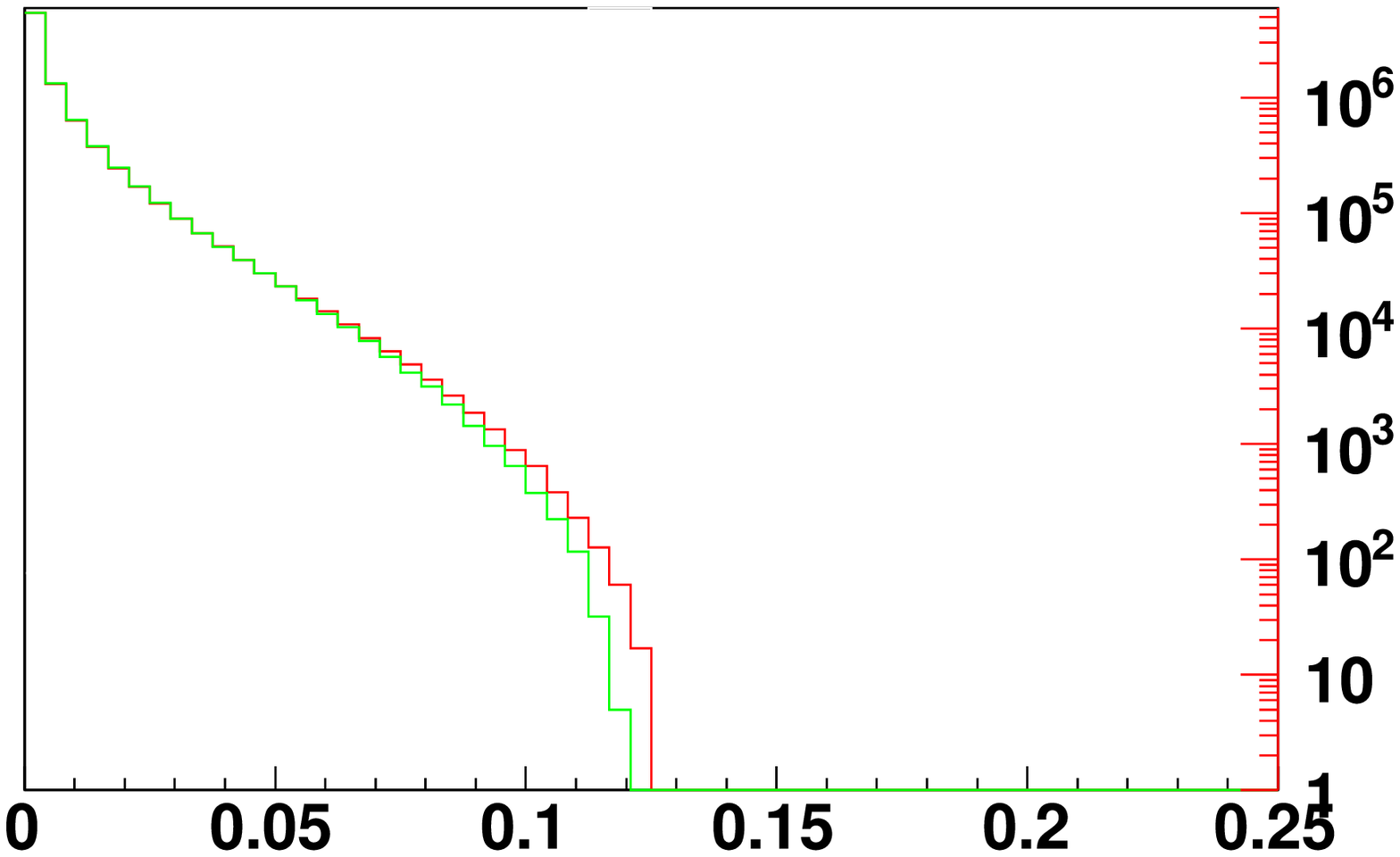}} &
\subfigure[$M_{\pi^+\bar\nu_ee^-}^2$]{\includegraphics[%
   width=0.48\columnwidth]{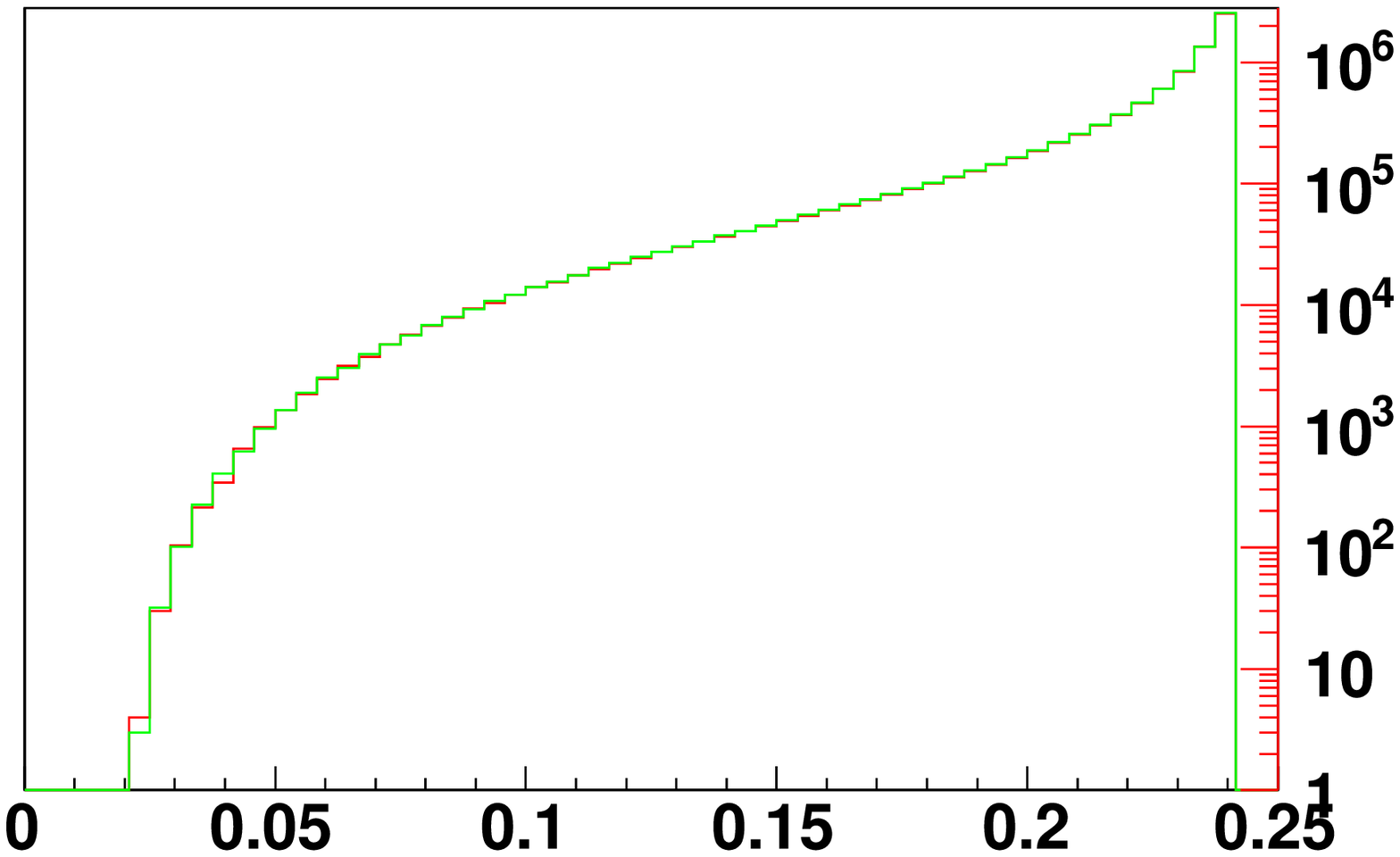}}
\end{tabular}
\caption{Distributions of scalar Lorentz invariants, in GeV$^2$
(GeV$^2/c^4$, $c=1$) units,  constructed from the decay
products in  $K^0\to \pi^+ e^- \bar\nu_e$ channel. The most  sensitive invariants to photon energy are plotted.
Single kinematical branch is used, thus the phase space is exact.
The
red (darker grey) line is standard {\tt PHOTOS}, the green is with exact
Matrix Element.
 The fraction of accepted bremsstrahlung events is
  (8.6398 $\pm$ 0.0029) \%  in standard {\tt PHOTOS} run
and  (8.6958 $\pm$ 0.0029) \% when the matrix element is used.  \label{figK0e1}}
\end{figure}

\begin{figure}[htp!]
\begin{tabular}{ccc}
\subfigure[$M_{\pi^+\gamma}^2$]{\includegraphics[%
  width=0.48\columnwidth]{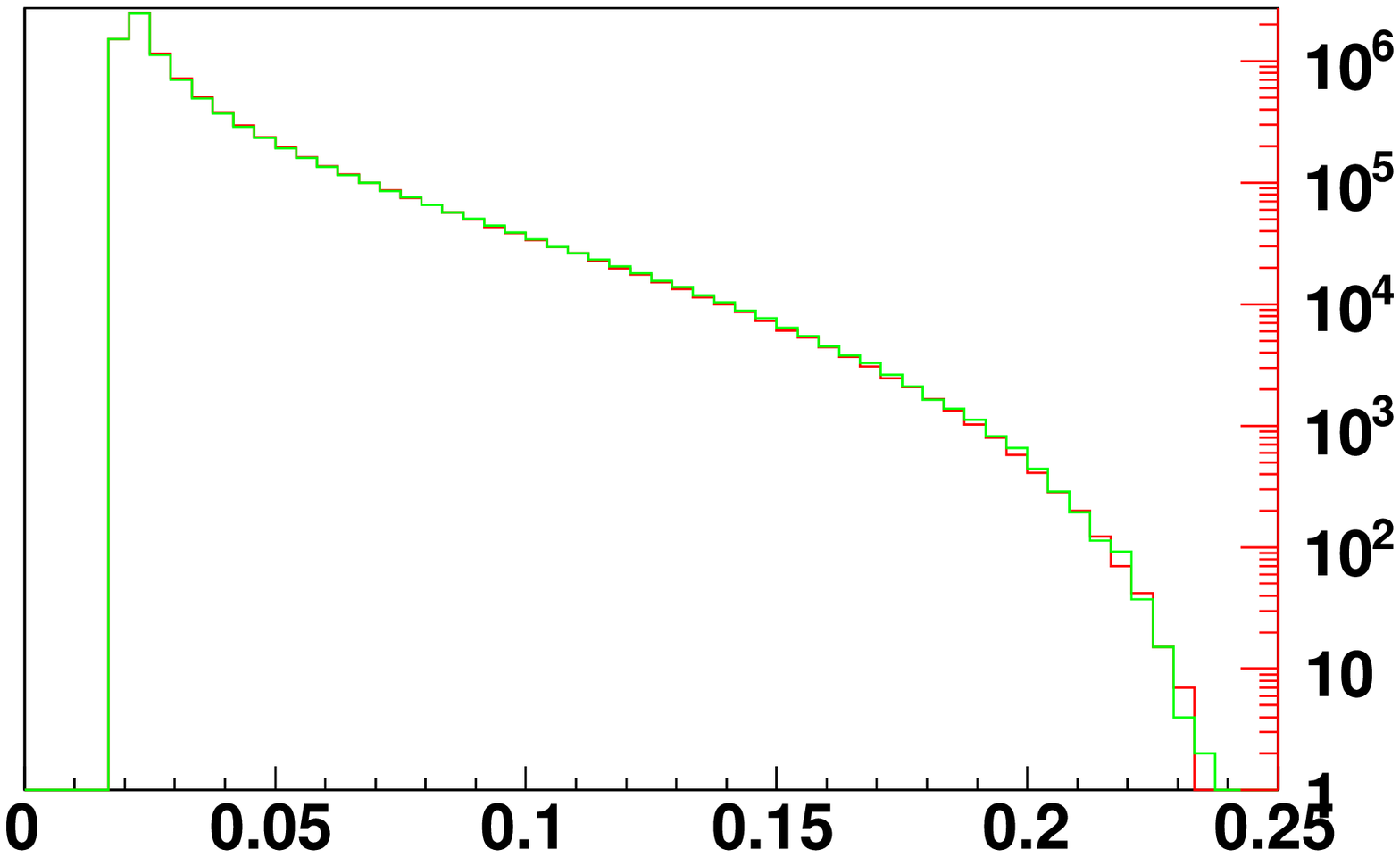}} &
\subfigure[$M_{e^-\gamma}^2$]{\includegraphics[%
  width=0.48\columnwidth]{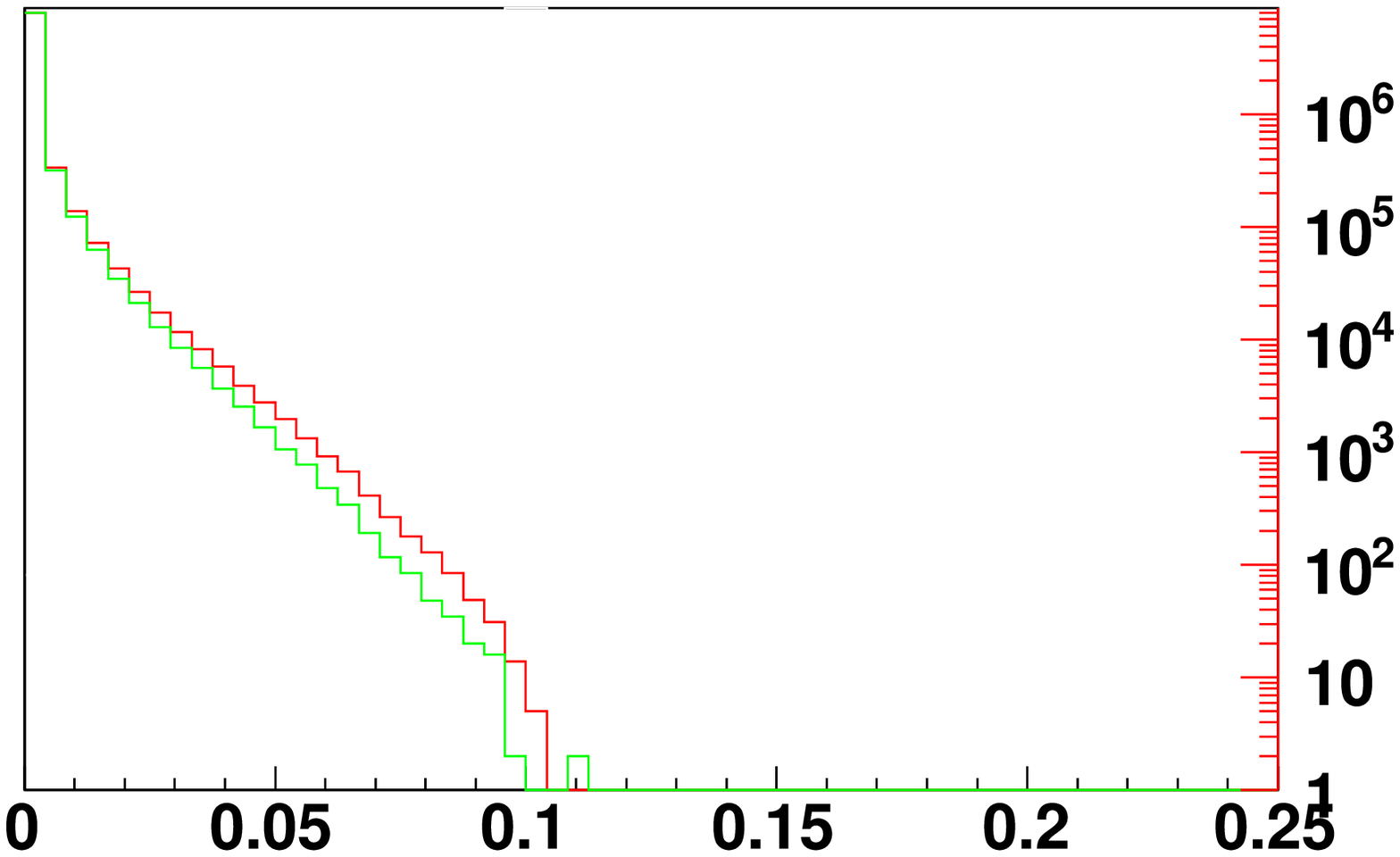}}
\\
\subfigure[$M_{\bar\nu_e\gamma}^2$]{\includegraphics[%
 width=0.48\columnwidth]{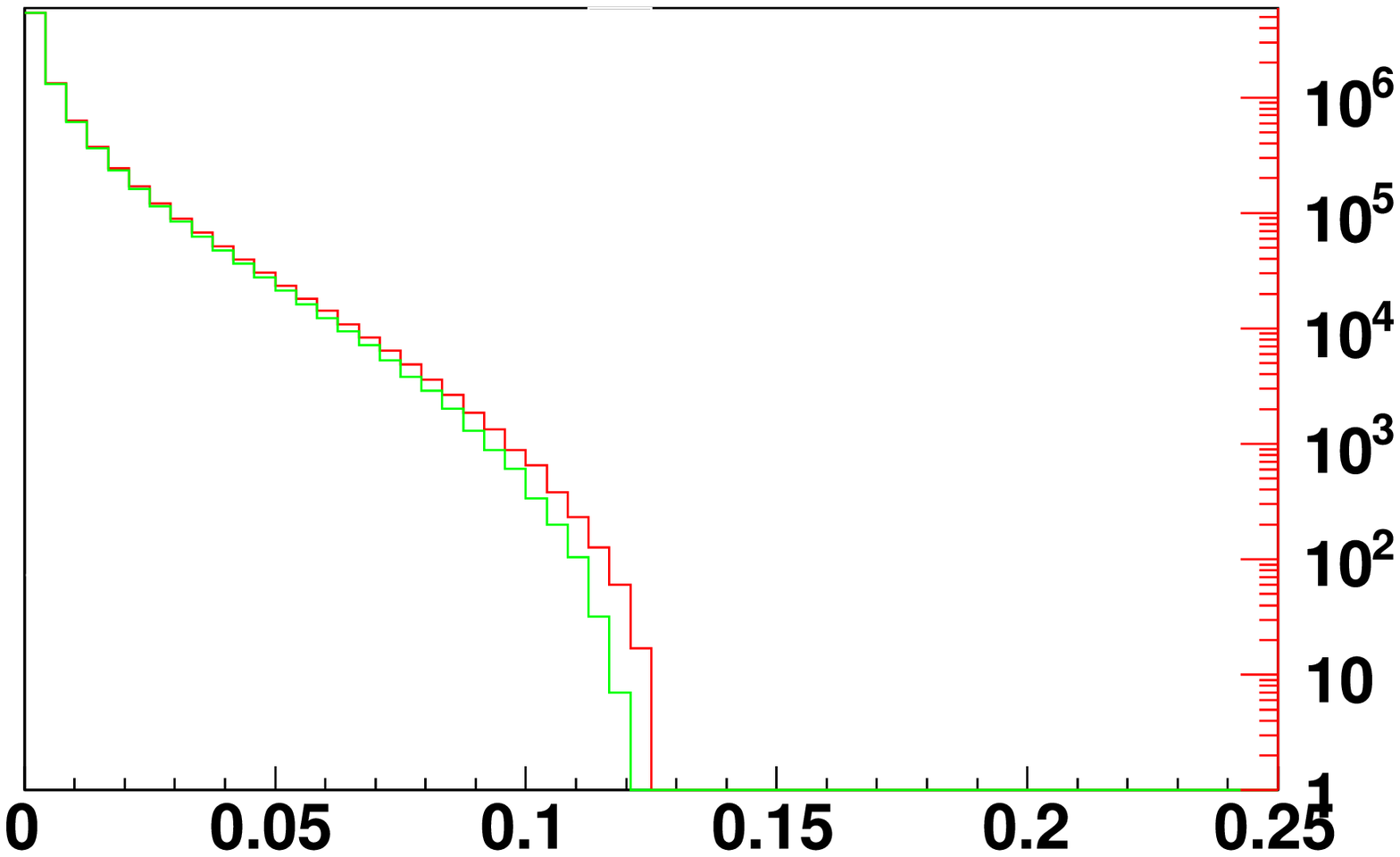}} &
\subfigure[$M_{\pi^+\bar\nu_ee^-}^2$]{\includegraphics[%
   width=0.48\columnwidth]{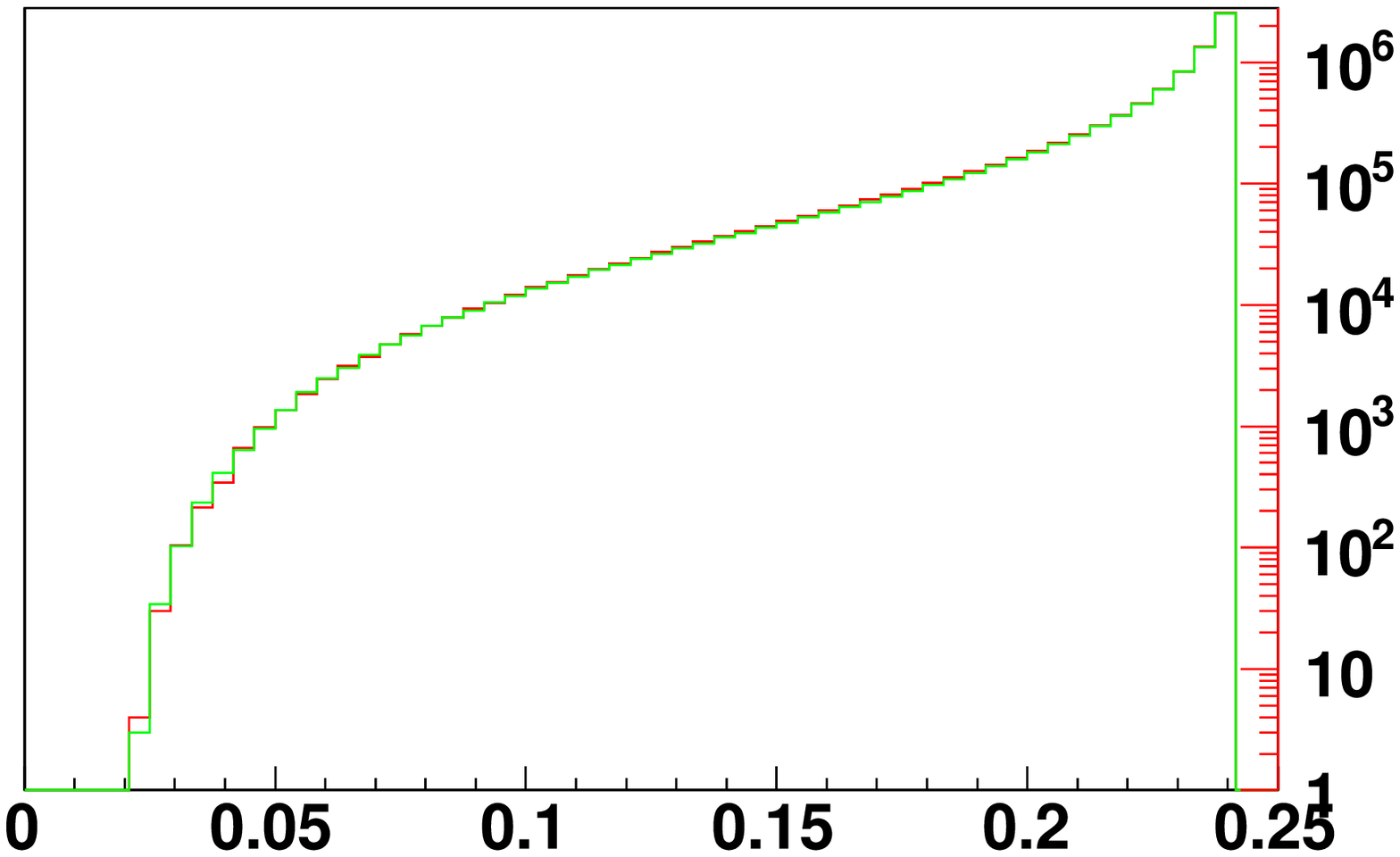}}
\end{tabular}
\caption{Distributions of scalar Lorentz invariants, in GeV$^2$
 (GeV$^2/c^4$, $c=1$) units,  constructed from the decay
products in  $K^0\to \pi^+ e^- \bar\nu_e$ channel. The most  sensitive invariants to photon energy are plotted.
Single kinematical branch is used, thus the phase space is exact.
The
red (darker grey) line is standard {\tt PHOTOS}, the green is with exact
Matrix Element.
 The fraction of accepted bremsstrahlung events is
  (8.6398 $\pm$ 0.0029) \%  in standard {\tt PHOTOS} run
and  (8.5235 $\pm$ 0.0029) \% when matrix element (\ref{ke3neutralnewC}) is used.
\label{figK0e2}}
\end{figure}

\begin{figure}[htp!]
\begin{tabular}{ccc}
\subfigure[$M_{\pi^+\gamma}^2$]{\includegraphics[%
  width=0.48\columnwidth]{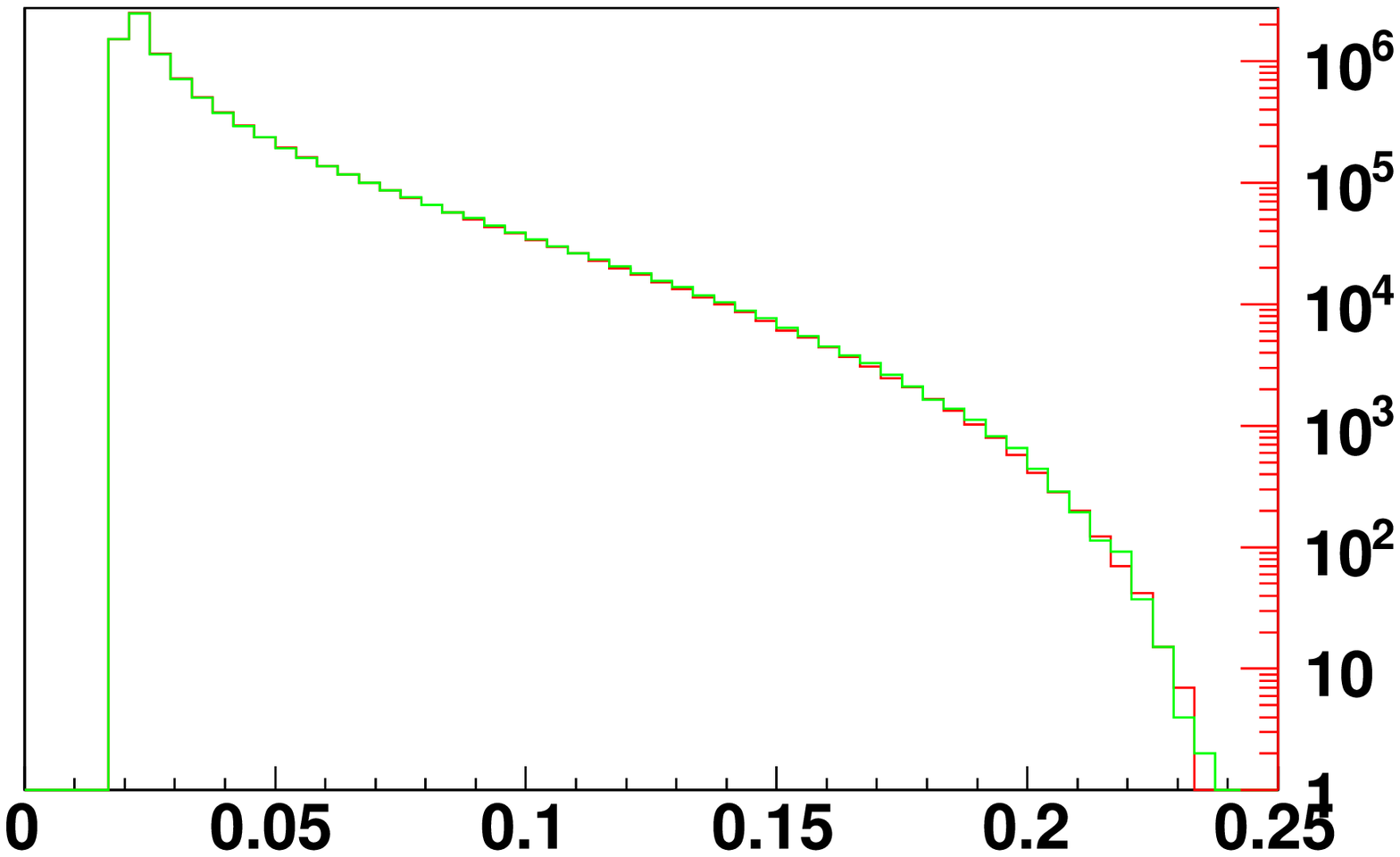}} &
\subfigure[$M_{e^-\gamma}^2$]{\includegraphics[%
  width=0.48\columnwidth]{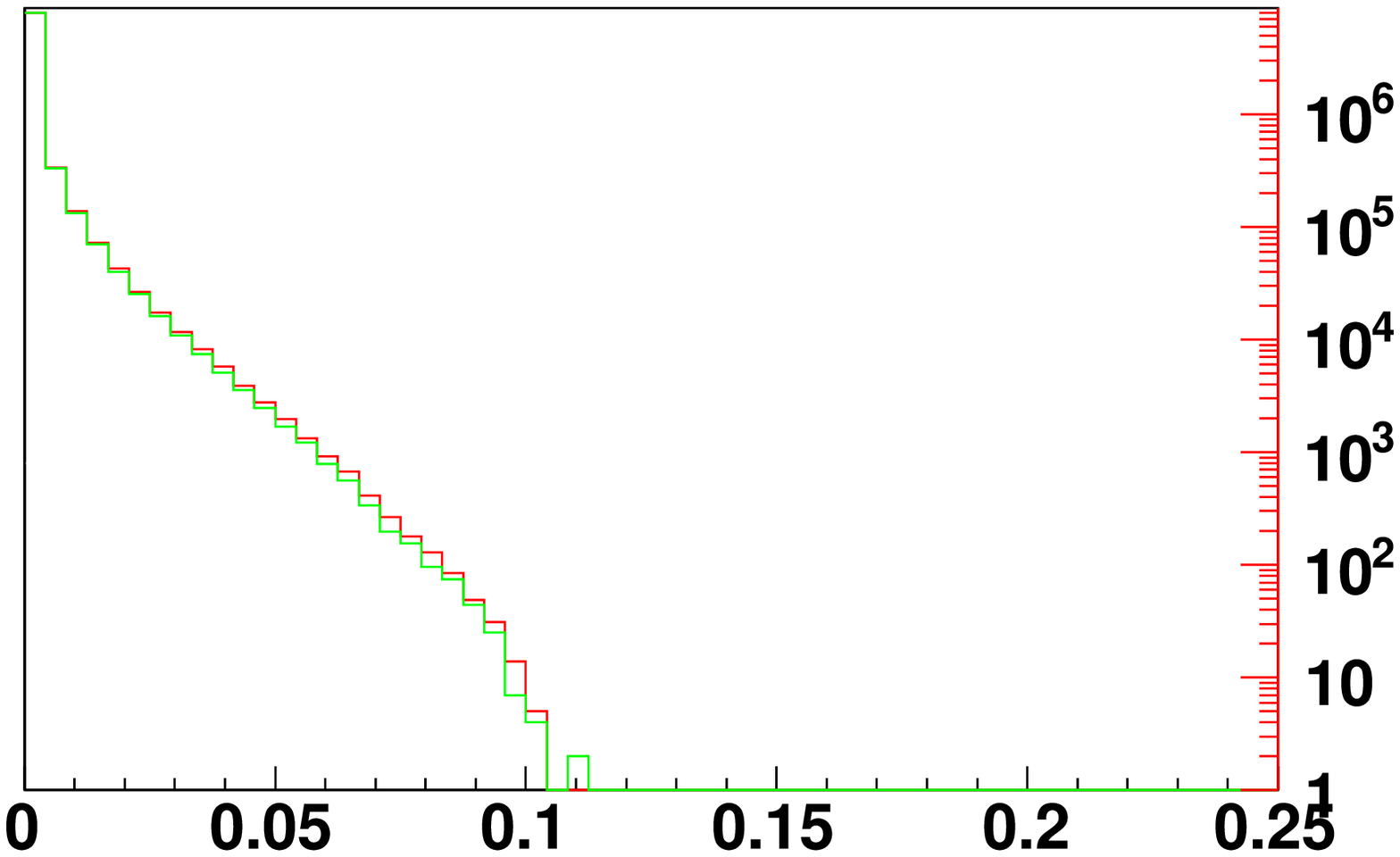}}
\\
\subfigure[$M_{\bar\nu_e\gamma}^2$]{\includegraphics[%
 width=0.48\columnwidth]{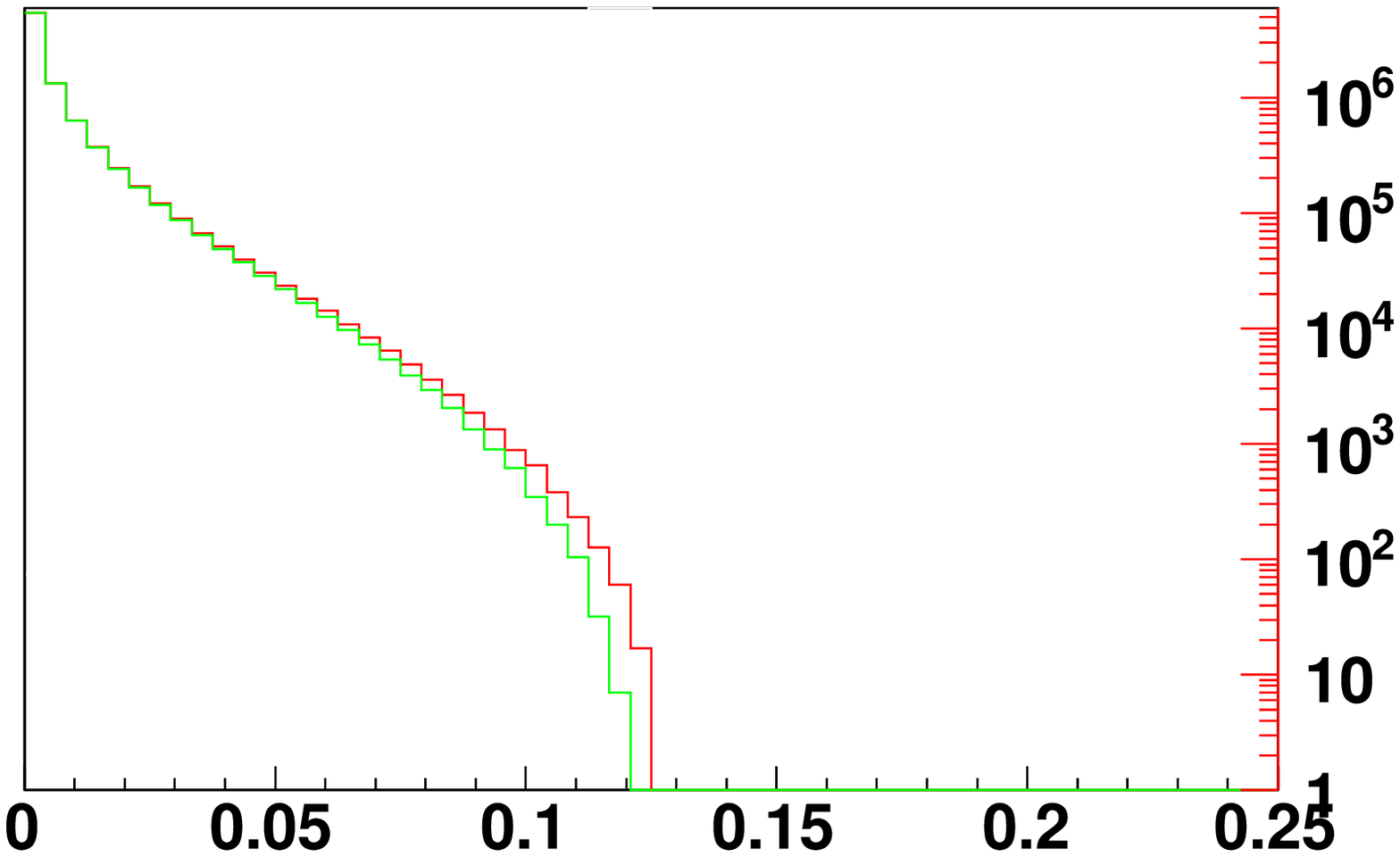}} &
\subfigure[$M_{\pi^+\bar\nu_e e^-}^2$]{\includegraphics[%
   width=0.48\columnwidth]{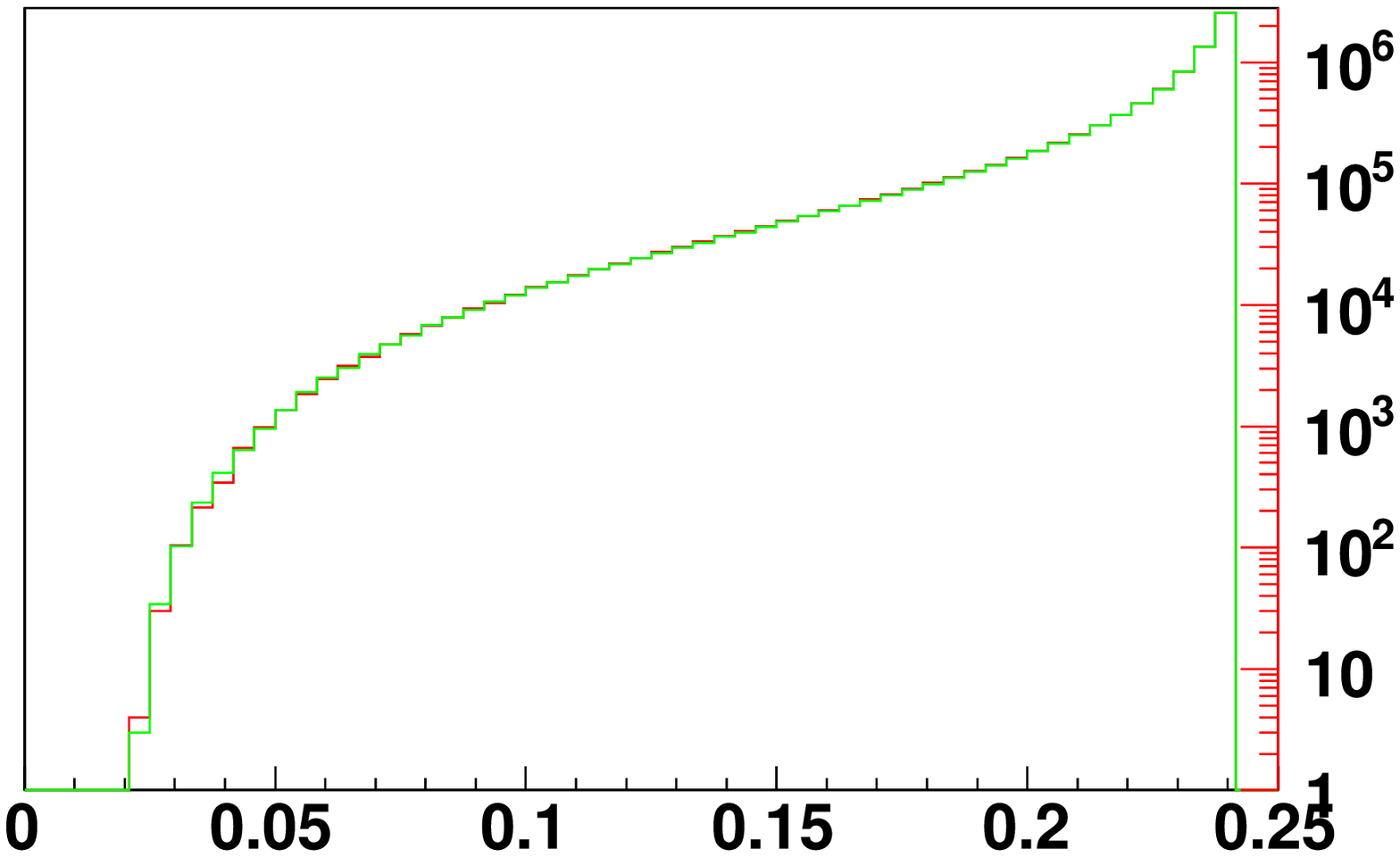}}
\end{tabular}
\caption{Distributions of scalar Lorentz invariants, in GeV$^2$
(GeV$^2/c^4$, $c=1$) units,  constructed from the decay
products in  $K^0\to \pi^+ e^- \bar\nu_e$ channel. The most  sensitive invariants to photon energy are plotted.
Single kinematical branch is used, thus the phase space is exact.
The
red (darker grey) line is standard {\tt PHOTOS}, the green is with exact
Matrix Element.
 The fraction of accepted bremsstrahlung events is
  (8.6398 $\pm$ 0.0029) \%  in standard {\tt PHOTOS} run
and (8.5928 $\pm$ 0.0029)  \%  when matrix element (\ref{ke3neutralnewC}) with
improvement of formula (\ref{ke3neutralnewD}) is used. \label{figK0e3}}
\end{figure}

\section{Summary}

In this study we have adopted the matrix element of
\cite{Cirigliano:2008wn} for the
emission kernel of {\tt PHOTOS} Monte Carlo
\cite{Barberio:1994qi,Golonka:2005pn,Nanava:2006vv}. We have
investigated  semileptonic decays of kaons into $\pi$ and lepton neutrino pair.
After modification for these channels,   {\tt PHOTOS} features exact
matrix element (three options)  and exact (or default) phase space. We  have
evaluated  the numerical size of the missing terms in publicly available
version of {\tt PHOTOS}. The difference is of the order of
0.2 \%, thus rather small, except for the distribution of lepton photon pair
invariant mass spectrum, where the difference is sizable at the high end
of the spectrum in case of  $K^0$ decay. 

On the technical side, we have also checked the algorithm.
We have compared
the case when the pre-sampler
is active for possible collinear singularity along lepton only
(for which the phase space is exact and phase space Jacobians are explicit), and the case
when both pre-samples for singularities
along lepton and charged $\pi$ directions are active.
Such studies for more than 2 body decays were not documented until now.
 We found the differences to be below
0.05 \%.
Our  work is supplemented with a larger set of figures, which are
 available from
the web-page \cite{web-Kl3}.

We conclude that for {\tt PHOTOS} version 2.15 or higher, and for
$K_{l3}$ decays, the precision level with respect to
matrix elements based simulation is 0.2\% or better.
This conclusion extends naturally to the
case of multiple photon emission.
We have identified  the factorization
properties of the matrix element, which were the reason why the differences were small.
On the other hand,
 our error estimation is not complete. It does not
include any discussion of uncertainty in the  matrix elements
 due to assumptions of the models used for their calculation. Also,
non leading effects of virtual corrections, which are expected
to contribute at a similar 0.2 \% level, are not discussed. They depend on the details
of hadronic interactions and can not be tackled in discussion
of  bremsstrahlung only.

\vskip 2 mm
\centerline{ \bf Acknowlegdements}
\vskip 2mm

We would like to thank Tomasz Przedzinski for help in obtaining the numerical results presented
in this paper and Brigitte Bloch-Devaux for help in improving the paper readability. We would like to thank referee of EPJC for idetifying  an error in our 
handling of formula (14) of Ref.~\cite{Cirigliano:2008wn}. 

\providecommand{\href}[2]{#2}
\end{document}